\documentclass[aps, pre, floatfix,
reprint,groupedaddress,
showpacs
]{revtex4-1}

\usepackage{graphicx}
\usepackage{bm}

\usepackage{amsmath}
\usepackage{amssymb}
\usepackage{color}
\usepackage{subfigure}
\usepackage{times}
\usepackage{float}
\usepackage{bm}
\usepackage[sort&compress]{natbib}

\usepackage[colorinlistoftodos,prependcaption,textsize=small]{todonotes}

\usepackage{hyperref}

\hypersetup{
  colorlinks=true,
citecolor=black,
linkcolor=black,
urlcolor=black
  }

\usepackage[T1]{fontenc}
\usepackage[utf8]{inputenc}

\newcommand{\bdt}[1]{{\color{blue} #1}}

\renewcommand{\bdt}[1]{{\color{black} #1}}

\begin{document}

\title{Trapped by the drift}

\author{Piotr Kubala}
  \email{piotr.kubala@doctoral.uj.edu.pl}
\author{Micha\l{} Cie\'sla}
  \email{michal.ciesla@uj.edu.pl}
\author{Bart\l{}omiej Dybiec}
  \email{bartek@th.if.uj.edu.pl}

  \affiliation{Institute of Theoretical Physics, Department of Statistical Physics, Jagiellonian University, \L{}ojasiewicza 11, 30-348 Krak\'ow, Poland}


\date{\today}

\begin{abstract}
The diffusion type is determined not only by microscopic dynamics but also by the environment properties. For example, the environment's fractal structure is responsible for the emergence of subdiffusive scaling of the mean square displacement in Markovian systems because the presence of non-trivially placed obstacles puts constraints on possible displacements. We investigate how the additional action of drift changes properties of the diffusion in the crowded environment. It is shown that the action of a constant drift increases chances of trapping, which suppresses the persistent ballistic motion. Such a diffusion becomes anisotropic because the drift introduces a preferred direction of motion which is further altered by interactions with obstacles. Moreover, individual trajectories display a high level of variability, which is responsible for the macroscopic properties of the diffusing front.
\bdt{Overall, the interplay between drift, diffusion and crowded environment, as measured by the time-averaged mean square displacement, is responsible for the emergence of superdiffusive and subdiffusive patterns in the very same system. Importantly, in contrast to free motion, the constant drift can enhance signatures of subdiffusive motion.}

\end{abstract}


%
\maketitle

\section{Introduction}

Diffusion \cite{metzler2000,metzler2004} is one of the fundamental processes frequently encountered in everyday life. From the macroscopic point of view, but not limited to, it is responsible for the equilibration of concentrations. Microscopically, the diffusion is produced by constant collisions between particles \cite{coffey2012langevin}. The macroscopic and microscopic images of diffusion can be combined in the stochastic description \cite{risken1996fokker,berkowitz2002physical,scher2002towards}. On the one hand, such an approach allows for inspection of properties of individual trajectories, while, on the other hand, it provides a possibility of calculation of probability densities from an ensemble of trajectories \cite{janicki1994}. These frameworks are applied both on  experimental, e.g., single-particle tracking \cite{jeon2011} or concentration measurement \cite{dguo2008molecular}, and theoretical levels \cite{cherstvy2013anomalous}.

Brownian motion \cite{brown1828,borodin2002} is a classic example of Markovian diffusion \cite{gardiner2009}. For such a diffusion, the variance of the particle position grows linearly in time. From the mathematical perspective, if a particle's displacements are independent and identically distributed according to a density with the finite variance, the particle position follows the normal (Gaussian) distribution with the linearly growing variance. Due to the central limit theorem \cite{feller1968}, the Gaussian distribution is the attractor in the probability space and, as such, is frequently recorded in the plenitude of diffusive systems. Nevertheless, experimental observations show numerous deviations from the linear scaling of variance. These deviations emerge because of violations of assumptions regarding the independence or the boundedness of displacements. They can also be produced by the structure of the environment in which the diffusion occurs \cite{saxton1994anomalous}.
Such situations, in contrast to normal diffusion, are referred to as anomalous diffusion \cite{metzler2000}.

\bdt{The properties of diffusion are determined by the characteristics of the environment in which diffusion occurs.
The structure of the medium plays an especially important role in anomalous transport in heterogeneous media like porous and fractured geological formations
\cite{horbach2017anomalous,edery2014origins,scher2002towards,di2003anomalous,berkowitz2002physical} or biological \cite{tolic2004anomalous,leijnse2012diffusion} realms.
Properties of the medium along with external forcing determines the performance of electrophoresis and chromatography.
The interaction with obstacles can result, among others, in non-Fickian diffusion of alcohol in silica \cite{zhokh2017application},
apparent anomalous diffusion in the cytoplasm \cite{kalwarczyk2017apparent}.
Structure of the environment can be also used  for particle separation or filtering \cite{fatin2004size}.
}

Traditionally, the diffusion type is quantified using the mean square displacement (MSD), see \cite{metzler2000,metzler2004}. The mean square displacement can be calculated from the set of $N$ trajectories $\{\bm{r}_i(t)\}$
\begin{equation}
    \langle \Delta \bm{r}^{\, 2} (t) \rangle = \frac{1}{N}\sum_{i=1}^N[\bm{r}_i(t)-\bm{r}(0)]^2.
    \label{eq:emsd}
\end{equation}
Traditionally, the diffusion type is discriminated by the exponent characterizing the growth of the MSD
\begin{equation}
 \langle \Delta \bm{r}^{\, 2} (t) \rangle = D_\alpha t^\alpha \propto t^\alpha.
 \label{eq:alpha}
\end{equation}
For $\alpha<1$, typically, the subdiffusion is recorded, while for $\alpha>1$ the superdiffusion is observed. The limiting case of $\alpha=1$ corresponds to the standard diffusion, while $\alpha=2$ to the ballistic motion. The discrimination of the diffusion type solely on the mean square displacement can be misleading \cite{dybiec2009h}. Alternatively to the ensemble averaging, it is possible to calculate the time-averaged mean square displacement (TAMSD) using a single trajectory, see \cite{lubelski2008b,he2008,sokolov2012},
\begin{equation}
    \overline{ \delta^2(\Delta ,T)  } = \frac{1}{T-\Delta} \int_0^{T-\Delta} [ \mathbf{r}(t+\Delta) - \mathbf{r}(t)]^2 dt,\\
    \label{eq:tamsd}
\end{equation}
where $T$ is the trajectory length. The relation between mean square displacement~(\ref{eq:emsd}) and time-averaged mean square displacement~(\ref{eq:tamsd}) is discussed in Appendix~\ref{sec:timeaverages}.

Typically, subdiffusion is observed in crowded \cite{sokolov2012}, gel-like, or fractal \cite{bouchaud1990} environments as it can be attributed to trapping events \cite{weeks2002subdiffusion}. Therefore, it is frequently recorded in a cellular realm \cite{sokolov2012,tolic2004anomalous,leijnse2012diffusion}, nevertheless the active superdiffusion \cite{caspi2002}, which can be followed by subdiffusion or normal diffusion can be also observed. Superdiffusion can also be recorded in rotating flows \cite{solomon1993} or optical systems \cite{barthelemy2008}.
Observations of large-scale displacements of spider monkey \cite{ramosfernandez2004}, jackals \cite{atkinson2002}  or albatrosses \cite{viswanathan1996,edwards2007} also show superdiffusion.

\bdt{The main aim of the current research is to explore properties of diffusion in the crowded environment \cite{sokolov2012} under the action of the constant drift. The combined action of the deterministic force and obstacles can result in nontrivial properties of the effective motion.} Whereas the drift is responsible for the emergence of the ballistic diffusion, i.e., quadratic (superlinear) scaling of the mean square displacement, the obstacles limit potential displacements, and might significantly slow down the diffusive motion, even causing subdiffusion \cite{ciesla2014tracer, ciesla2015taming}. Furthermore, the presence of obstacles makes the environment heterogeneous, what in turn might result in substantial differences among individual trajectories.
Therefore, within the current research we focus on the examination of how the combined action of the deterministic force and obstacles changes properties of diffusion.


The model under study is described in the next section (Sec.~\ref{sec:model} Model).
Sec.~\ref{sec:results} (Results) analyses properties of diffusion in the crowded environment under the action  of the constant drift.
The manuscript is closed with Summary and Conclusions (Sec.~\ref{sec:summary}).
Additional information is moved to Appendices \ref{sec:langevin} and~\ref{sec:timeaverages}.
\bdt{Finally, the extra Appendix~\ref{sec:spherocylinders} further explores the role played by the shape of obstacles.}

%
%
\section{Model\label{sec:model}}

\begin{figure}[!h]
    \centering
    \includegraphics[width=0.8\columnwidth]{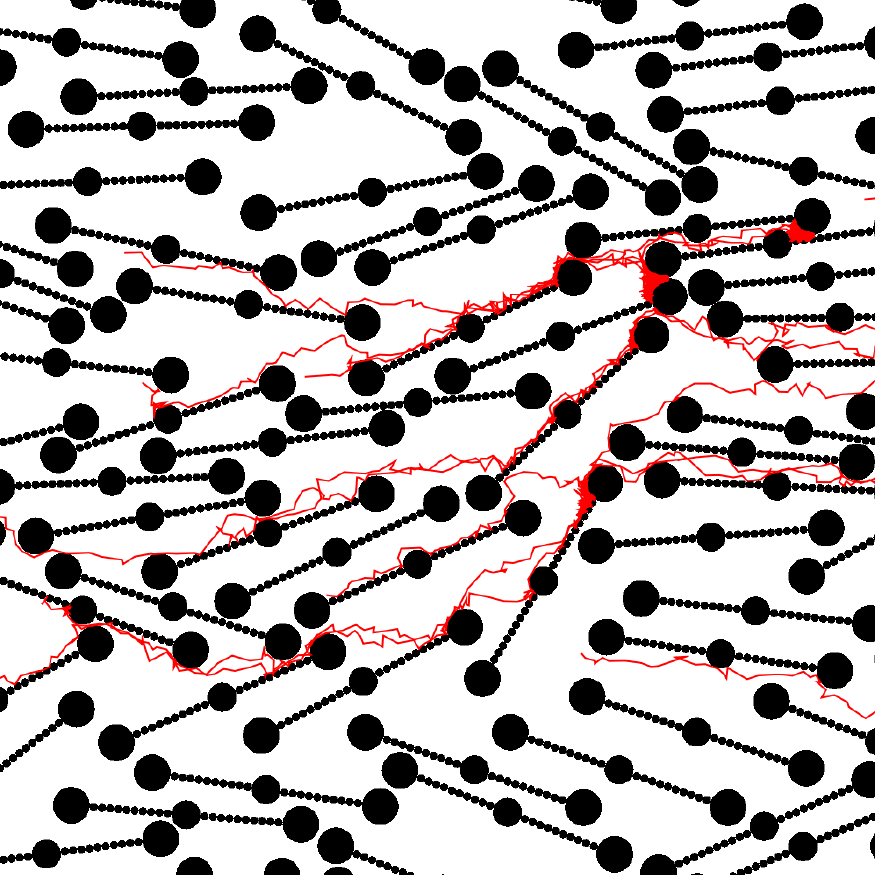}
    \caption{The 2D crowded environment filled with fibrinogen particles, which are built from three larger and twenty small balls.
    The standard deviation of the orientational angle $\theta$ \bdt{(angle between the drift direction and long axis of the molecule)} is set to $\sigma(\theta)=20^\circ$.
    Additional red curves depict  a few sample trajectories with the deterministic drift set to $\Lambda = 1$.}
    \label{fig:environment}
\end{figure}

The diffusion occurs in the two-dimensional space filled with randomly placed obstacles, which influence properties of motion. As obstacles, we use a model of fibrinogen particles \cite{Adamczyk2010} randomly placed on a two-dimensional surface. Each fibrinogen particle is built by three larger and twenty ($2\times 10$) smaller balls. Similar spatial patterns were also used in previous studies on effective driftless diffusion in two-dimensional crowded environments \cite{Ciesla2014, ciesla2015taming}. Sample, crowded, two-dimensional environment is presented in Fig.~\ref{fig:environment}. Additionally, in Fig.~\ref{fig:environment} a few sample trajectories are plotted in red.
\bdt{Nevertheless, for testing purposes we have repeated all simulations with fibrinogen molecules replaced by spherocylinders, see Appendix~\ref{sec:spherocylinders}.
These simulations have shown that absence of concave parts of obstacles does not affect the qualitative predictions of the model based on the fibrinogen.
Therefore, the environment built by fibrinogen molecules is considered as the main setup.
}

Contrary to the purely diffusive motion studied in \cite{ciesla2015taming}, here, we assume biased diffusion.
More precisely, we study the two-dimensional overdamped motion under the action of the Gaussian white noise and constant drift.
Changes in the particle (tracer) position $\bm{r}$ are described by the overdamped Langevin equation
\begin{equation}
    \frac{d \bm{r}}{dt}=\bm{\Lambda}+ \bm{\xi}(t),
    \label{eq:langevin2d}
\end{equation}
where $\bm{\Lambda}$ is the constant drift and $\bm{\xi}(t)$ is the two-dimensional Gaussian white noise \cite{gardiner1983,samorodnitsky1994}.
We assume that the noise is spherically symmetric, i.e., $\bm{\xi}(t)=(\xi_x(t),\xi_y(y))$, where  $\xi_x(t)$ and $\xi_y(t)$ are two independent one-dimensional Gaussian white noises satisfying
\begin{equation}
    \langle \xi_i(t) \rangle=0
\end{equation}
and
\begin{equation}
    \langle \xi_i(t)\xi_j(s) \rangle=\delta_{ij}\sigma_0^2 \delta(t-s),
\end{equation}
where $\delta_{ij}$ and $\delta(\cdot)$ are Kronecker and Dirac deltas, respectively.
The Langevin equation~(\ref{eq:langevin2d}) has the following discretized version
\begin{equation}
    \bm{r}_{i+1} = \bm{r}_{i} + \bm{\Lambda}\Delta t + \sigma_0 \bm{\xi}_t \sqrt{\Delta t},
    \label{eq:langevin-discrete}
\end{equation}
where $\bm{\xi}_t$ is the sequence of independent two-dimensional, spherically symmetric normal random variables, and $\Delta t$ is the integration time step.
For more details on the difference between free Brownian motion and the motion under the action of the constant drift, see Appendix~\ref{sec:langevin}.
The Langevin equation~(\ref{eq:langevin2d}) describes the motion of a particle in the unrestricted environment.
Since the particle moves freely  between obstacles, see Fig.~\ref{fig:environment}, it is necessary to define the way how a particle interacts with obstacles.
We assume that interactions are of the excluded volume type. Therefore, the displacements overlapping fibrinogen particles are rejected.
The whole algorithm describing the microscopic dynamics can be summarized in the following way:
\begin{enumerate}
 \itemsep0em
    \item Pick a random starting position $\bm{r}_0$.
    \item Set time to 0, i.e., $t=0$.
    \item For each random step:
    \begin{enumerate}
    \itemsep0em
        \item calculate the displacement $\Delta \bm{r}$
        $$        \Delta \bm{r}=\bm{\Lambda}\Delta t + \sigma_0 \bm{\xi}_t \sqrt{\Delta t}   ;     $$
        \item if the line connecting $\bm{r}_i$ and $\bm{r}_{i} + \Delta \bm{r}$ does not overlap with any obstacle, accept the move, i.e., set $\bm{r}_{i+1}$ to $\bm{r}_{i+1}=\bm{r}_{i} + \Delta \bm{r}$;
        \item if the line connecting $\bm{r}_i$ and $\bm{r}_{i} + \Delta \bm{r} $ overlaps with any obstacle, reject the move, i.e., $\bm{r}_{i+1}=\bm{r}_{i}$;
        \item increase time $t$ by $\Delta t$, i.e., $t \to t + \Delta t$.
    \end{enumerate}

    \item{Repeat step 3 as long as $t< T$.}

\end{enumerate}

The constant  drift $\bm{\Lambda}$ ($\Lambda=|\bm{\Lambda}|$) acts along the $x$-axis, while the
obstacles, see Fig.~\ref{fig:environment}, are placed using the random sequential adsorption algorithm (RSA) \cite{Feder1980, Evans1993} on the $L \times L$  square with periodic boundary conditions.
To place a fibrinogen particle, it is necessary to generate a position of its center  and the orientational angle $\theta$ between the long axis of the fibrinogen particle and the $x$-axis.
The particle position is uniformly distributed on the $L \times L$ square. As we want to study how the orientational order of obstacles affects diffusion, the order is introduced by sampling the orientations of obstacles from the normal distribution $N(0^\circ,\sigma(\theta))$, whose center is $0^\circ$, so the preferred direction is horizontal, and $\sigma(\theta)$ is the standard deviation of the angle $\theta$ between the $x$-axis and the long axis of the molecule measured in degrees, i.e., $\theta \sim N(0^\circ,\sigma(\theta))$.
\bdt{The standard deviation of the Gaussian distribution controls the ordering of the resulting crowded environment.
With increasing $\sigma$, the environment becomes more disordered.
}
\bdt{Sequentially added fibrinogen particles are placed in such a way that they do not overlap.
More precisely, before adding a new fibrinogen molecule it is verified if it does not cross other molecules.
In the case of overlapping, new position and orientation are generated.
}
Adding of obstacles is stopped at the density of $\varrho=0.27$, i.e., when the fraction of the full space filled by fibrinogen is equal to $27\%$.
\bdt{The studied setup, especially under the action of constant drift resembles the Galton board \cite{gaspard1995chaotic,gaspard1989scattering,judd2007galton,arai2012randomness,barra2009fractality} as diffusing particles can travel around tunnels among molecules.
This similarity is especially well visible for highly ordered setups where molecules of the environment are parallel.
}
The simulation is performed on the image of a generated packing, therefore the natural length unit is one pixel.
\bdt{Each fibrinogen particle is built by three larger and twenty ($2\times 10$) smaller balls.}
The length of the single fibrinogen particle corresponded to $300$ pixels, \bdt{while radii of small and larger balls are respectively equal to 9, 33 and 41 pixels},
and the linear size of the packing is $L=3000$ pixels.
The noise intensity $\sigma_0$ is set to $\sigma_0=3/\sqrt{2}$.
The trajectories were simulated up to $T=10^5$.
\bdt{Typically, for each set of parameters, we have generated not-less than $10^5$ individual trajectories.}
The simulation length $T$ was imposed by the integration time step and computational resources.

The adjusting of the appropriate integration time step $\Delta t$ is crucial to obtain the correct results.
Therefore, we plotted the mean square displacement against time $t$ for a series of decreasing values of $\Delta t$ and chose the biggest one for which the mean square displacement already stabilized.
Therefore, the integration time step was adjusted by the self-consistency test. We used $\Delta t = 1$ for $\Lambda = 0$ and $\Delta t = 0.1$ for $\Lambda = 1$.
From the multiple trajectories (up to $10^5$) generated by Eq.~(\ref{eq:langevin-discrete}) it is possible to calculate the MSD, probability densities and other system characteristics.

The detailed analysis of the diffusion under constant drift in the crowded environment is presented in the next section.
In particular, we explore properties of diffusive patterns, mean square displacements, time-averaged mean square displacements, and related characteristics.

%
%
\section{Results and discussion\label{sec:results}}

Diffusive patterns are affected by the action of the drift and the presence of obstacles, see Fig.~\ref{fig:environment}.
In short, \bdt{in the absence of obstacles}, both in the presence and the absence of the drift, the Gaussian profile of the probability density function of finding a tracer (particle), symmetrically widens.
Differences are recorded in the behavior of the modal value.
In the absence of the drift the mode does not move, i.e., it stays at the starting point.
Under the action of the drift the mode moves with a constant velocity along the drift, resulting in the ballistic motion.
Therefore, the presence of drift hinders the problem of quantifying the diffusion type as the mean square displacement, due to the drift, grows quadratically, i.e., ballistically, in time, see Eqs.~(\ref{eq:msd1d}) and~(\ref{eq:msd2d}).
Therefore, in order to discriminate the random component of the motion type one can subtract the constant trend, i.e., to use the variance instead of the mean square displacement, see Eqs.~(\ref{eq:std1d}) and~(\ref{eq:std2d}).
The situation significantly changes when the diffusion occurs in the crowded environment.
Obstacles confine the motion of particles, see red trajectories in Fig.~\ref{fig:environment}, consequently, the diffusion is no longer fully free.
A particle motion can follow the drift for a short time only.
After some time, it has to interact with obstacles and diffusion starts to differ from the free diffusion.
Furthermore, due to space heterogeneity, individual paths differ very much between realizations.

\begin{figure}[!h]
    \centering
    \includegraphics[width=\columnwidth]{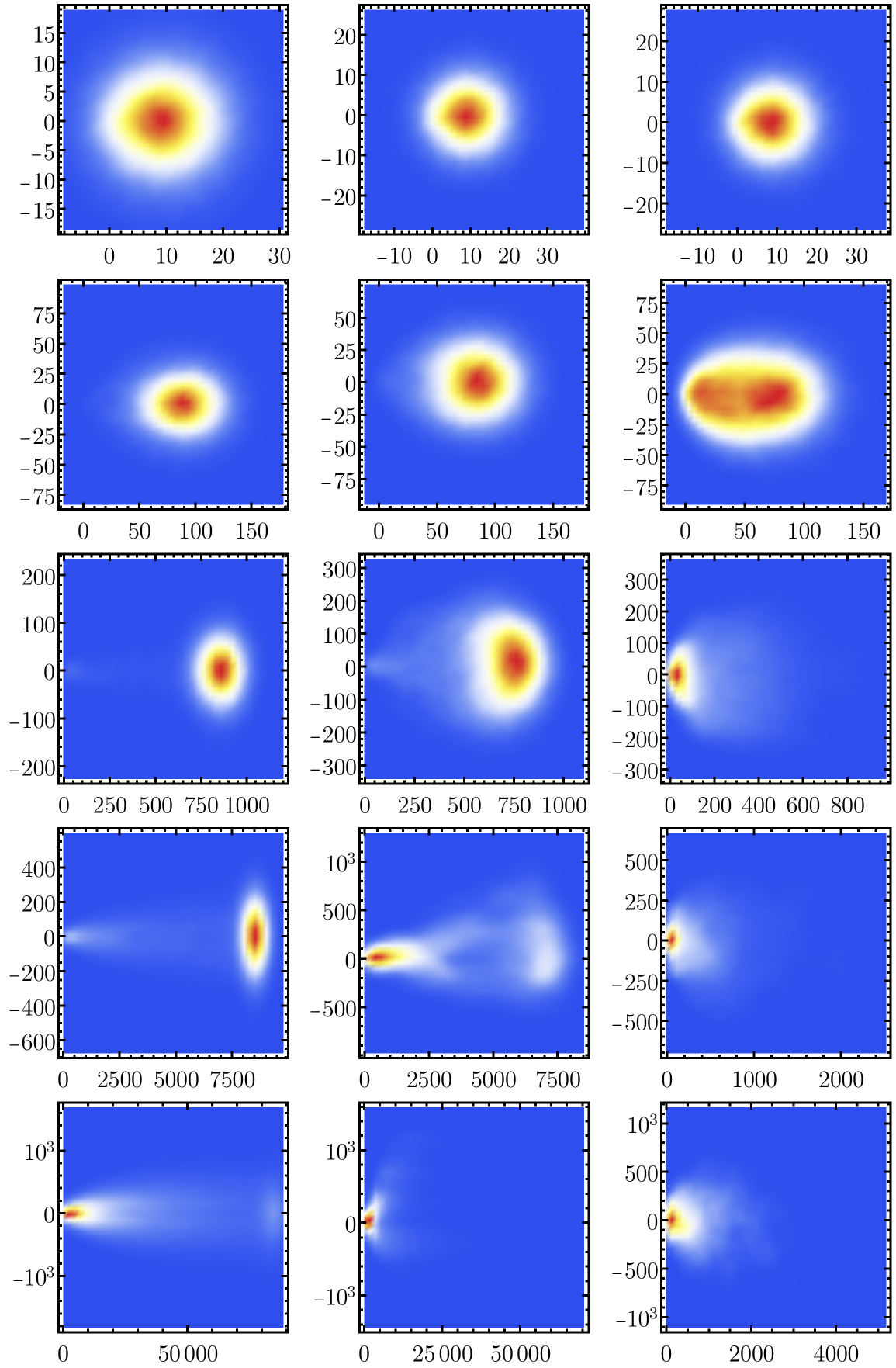}
    \caption{Probability densities $p(x,y)$ for the drift $\Lambda=1$ at various timestamps $t\in\{10^1,10^2,10^3,10^4,10^5\}$ (from top to bottom).
    Subsequent columns correspond to increasing values of the standard deviation $\sigma(\theta)$ of the orientational angle $\theta$, i.e., $\sigma(\theta)\in\{0^\circ,20^\circ,180^\circ\}$. Red, white, and blue colors correspond to high, mediocre and low density, respectively, however the scale is panel-dependent, i.e., the same colors in different panels can correspond to different values.
    }
    \label{fig:2dhis}
\end{figure}

\begin{figure}[!h]
    \centering
    \begin{tabular}{cc}
    \includegraphics[width=\columnwidth]{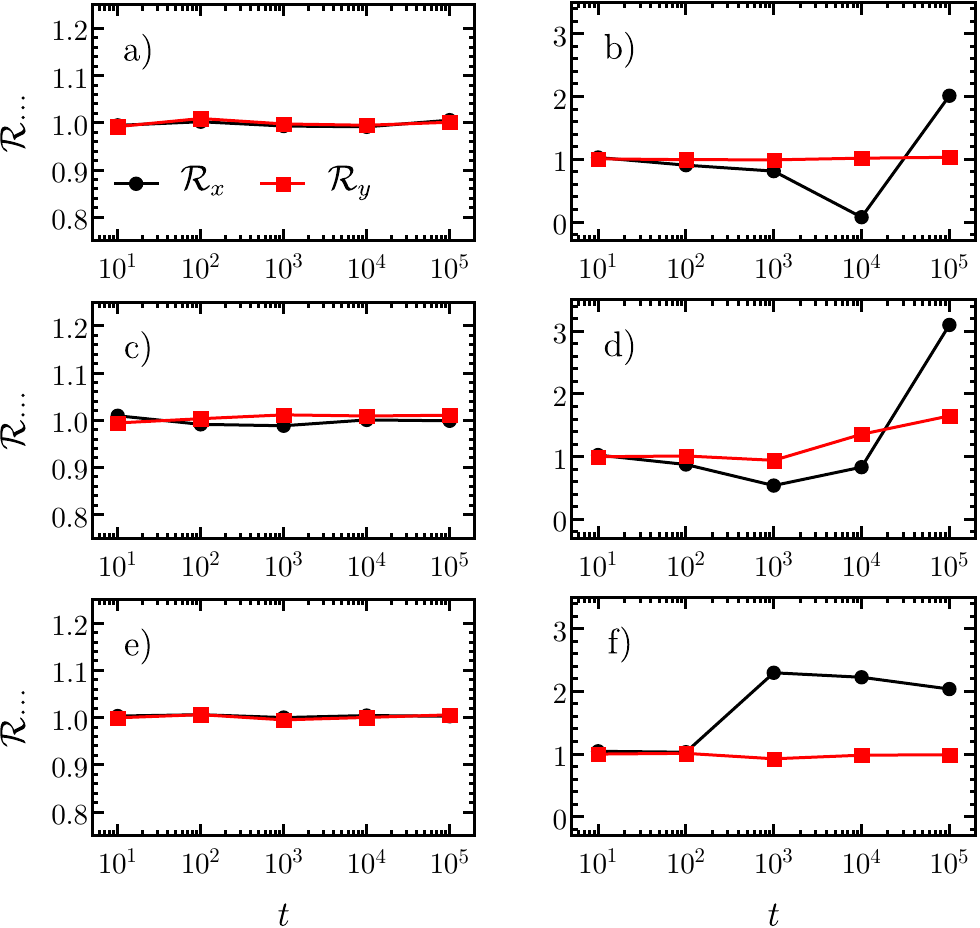}
    \end{tabular}
    \caption{
    Time dependence of width ratios $\mathcal{R}_x$ [black (darker) dots] and $\mathcal{R}_y$ [red (lighter) squares], see Eq.~(\ref{eq:width}), in the absence ($\Lambda=0$, left column) and presence ($\Lambda=1$, right column) of the drift.
    Various rows correspond to different values of the standard deviation $\sigma(\theta)$ of the orientational angle $\theta$: $\sigma(\theta)\in\{0^\circ,20^\circ,180^\circ\}$  (from top to bottom).
    The legend is included in the top left panel ($a$).
    Lines are drawn to guide the eye only.
    }
    \label{fig:widthratio}
\end{figure}

Fig.~\ref{fig:2dhis} shows probability densities of trajectories at displacements $\bm{r}_i(t)-\bm{r}_i(0)$ for $t\in\{10^1,10^2,10^3,10^4,10^5\}$ (from top to bottom), and the drift $\Lambda=1$, estimated from an ensemble of $10^5$ trajectories.
Subsequent columns (from left to right) correspond to the increasing standard deviation of the orientational angle $\sigma(\theta)$ ($\sigma(\theta)\in\{0^\circ,20^\circ,180^\circ\}$).
We do not show results corresponding to $\Lambda=0$ as in the driftless case the probability densities are almost symmetric, see the discussion below and the left column of Fig.~\ref{fig:widthratio}.
Histograms depicted in Fig.~\ref{fig:2dhis} show that the $p(x,y;t)$ densities are almost symmetric along the $y$-axis also for a non-zero drift.
At the same time, motion along the horizontal axis becomes more dispersed.
Fig.~\ref{fig:2dhis} clearly indicates that the drift can produce the particle's movement for a limited time only, which is determined by the presence of obstacles building traps.
The duration of the deterministic phase of the motion depends on the ordering, i.e., $\sigma(\theta)$.
More precisely, it is the decreasing function of the standard deviation $\sigma(\theta)$ of the orientational angle $\theta$.
During the motion, a substantial fraction of particles becomes trapped near the initial position.
A further part of particles gets trapped after the transient deterministic part of the motion.
Finally, some of the particles still move to the right shifting the hairy part of the probability density from the left hand side to the right.
The trapping is also visible on the single trajectory level, see Fig.~\ref{fig:environment}.
The trapping, which is already well visible for $\Lambda=1$, can be amplified by increasing the drift strength $\Lambda$ (results not shown), because increasing drift increases the likelihood of rejecting generated displacements.

Properties of the motion in the driftless case and under the action of the constant drift in the crowded environment are quantitatively studied using the following quantifiers
\begin{equation}
 \mathcal{R}_x=\frac{x_{0.9}-x_{0.5}}{x_{0.5}-x_{0.1}}\;\;\;\;\mbox{and}\;\;\;\;\mathcal{R}_y=\frac{y_{0.9}-y_{0.5}}{y_{0.5}-y_{0.1}},
\label{eq:width}
 \end{equation}
where $x_{\dots}$ and $y_{\dots}$ stand for quantiles of a given order, e.g., $x_q(t)$ is defined by
\begin{equation}
 q = \int_{-\infty}^{x_q(t)} p(x;t) dx.
 \label{eq:quantile}
\end{equation}
In the above equation $p(x;t)$ stands for the marginal, time-dependent probability distribution of $x$. Quantiles $y_{\dots}$ are defined in an analogous way. The ratio defined by Eq.~(\ref{eq:width}) measures the fraction of widths of intervals containing $40\%$ of probability mass below ($x_{0.5}-x_{0.1}$) and above ($x_{0.9}-x_{0.5}$) the median ($x_{0.5}$). Its value reflects the symmetry of the probability density: for $\mathcal{R}=1$, the intervals' widths in the numerator  and the denominator are the same. If $\mathcal{R}<1$, the probability density is skewed to the left, while for $\mathcal{R}>1$ to the right.

The time dependencies of width ratios $\mathcal{R}_x$ and $\mathcal{R}_y$ for diffusion without and with the drift are shown in Fig.~\ref{fig:widthratio}.
Subsequent rows correspond to various values of the standard deviation $\sigma(\theta)$ of the orientational angle $\theta$ (from top to bottom $\sigma(\theta)\in\{0^\circ,20^\circ,180^\circ\}$). The left column corresponds to the driftless case $\Lambda=0$, and the right one to $\Lambda=1$. In the absence of drift, the probability densities are almost symmetric, so the widths ratio defined in Eq.~(\ref{eq:width}) is close to $1$. Under the action of the drift, only the symmetry along the $y$-axis is still present, see Fig.~\ref{fig:2dhis} and the right column of Fig.~\ref{fig:widthratio} (red squares). Only one exception is visible for $\sigma(\theta)=20^\circ$, which is the artifact of finite size of the given environment. More precisely, in this particular realization, the placing of fibrinogens introduces a weak bias. At the same time, along the $x$-axis (black circles), action of the drift and presence of obstacles significantly changes the spread of the diffusive packet, see right column of Fig.~\ref{fig:widthratio}.
Horizontally, the packet becomes initially skewed to the left, i.e., the $x_{0.9}-x_{0.5}$ interval is significantly narrower than the $x_{0.5}-x_{0.1}$ interval but after some time, due to termination of the deterministic phase of the motion, it becomes skewed to the right, i.e., the $x_{0.9}-x_{0.5}$ interval becomes wider than the $x_{0.5}-x_{0.1}$.
Therefore, initially $\mathcal{R}_x$ is smaller than 1 as the diffusive front moves to the right, but after some critical ``front-trapping'' time it becomes positive because the diffusive front (significant fraction of particles) is trapped.
More precisely, after trapping the diffusive front, some of the particles still move to the right making the width ratio $\mathcal{R}_x$ larger than 1.
With increasing level of randomness, i.e., $\sigma(\theta)$, the duration of the deterministic phase of the motion decreases and $\mathcal{R}_x$ becomes larger than 1 faster, see black circles in the right panel of Fig.~\ref{fig:widthratio}.
These observations suggest that analyzing properties of diffusive processes in crowded environments, especially in the presence of drift, using the mean square displacement, see Eq.~(\ref{eq:emsd}), due to enormous differences between various trajectories, requires special care.  

\begin{figure}[!h]
    \centering
    \includegraphics[width=0.8\columnwidth]{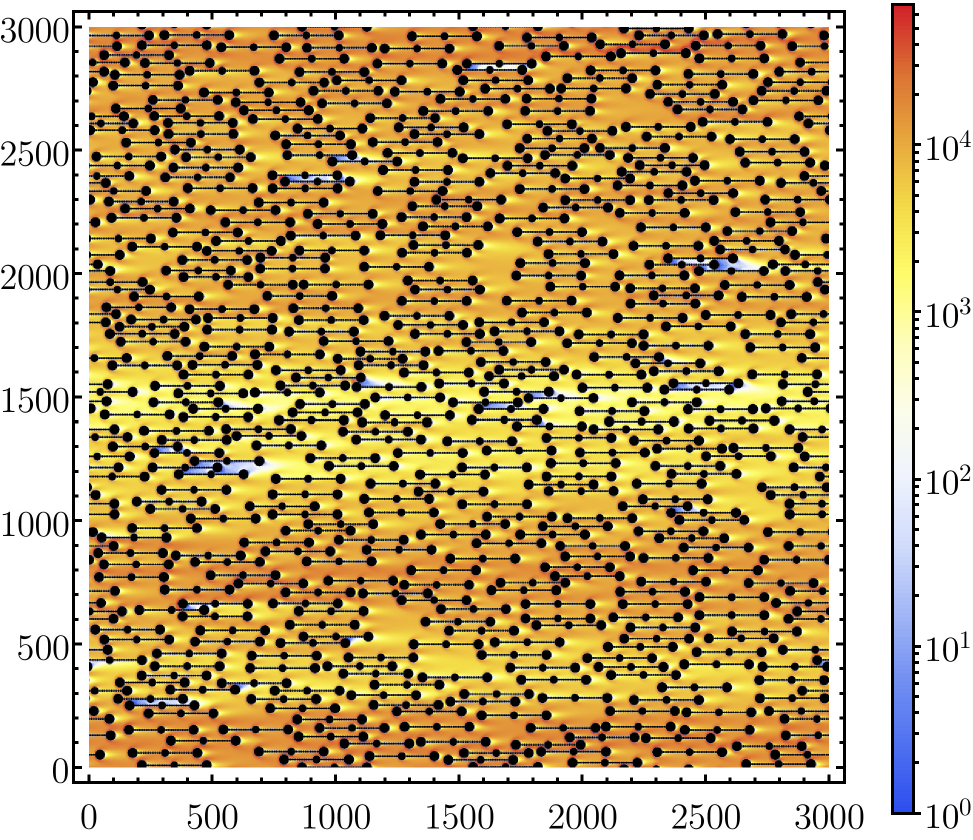} \\
    \vspace{8pt}
    \includegraphics[width=0.8\columnwidth]{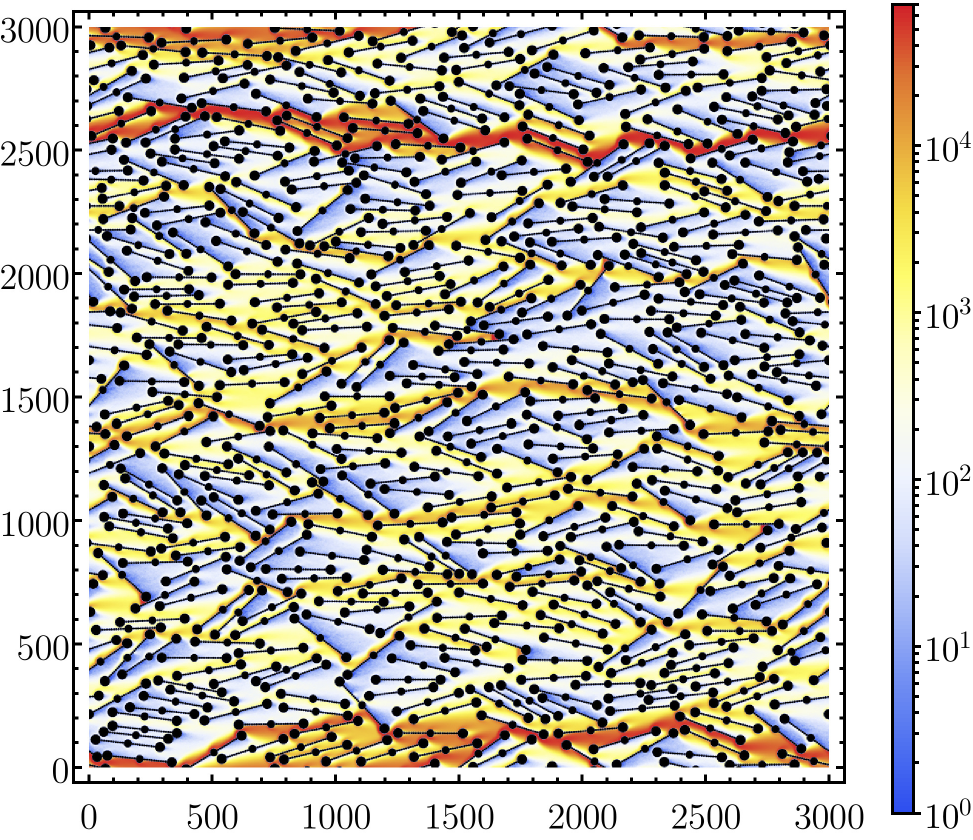}\\
    \vspace{8pt}
    \includegraphics[width=0.8\columnwidth]{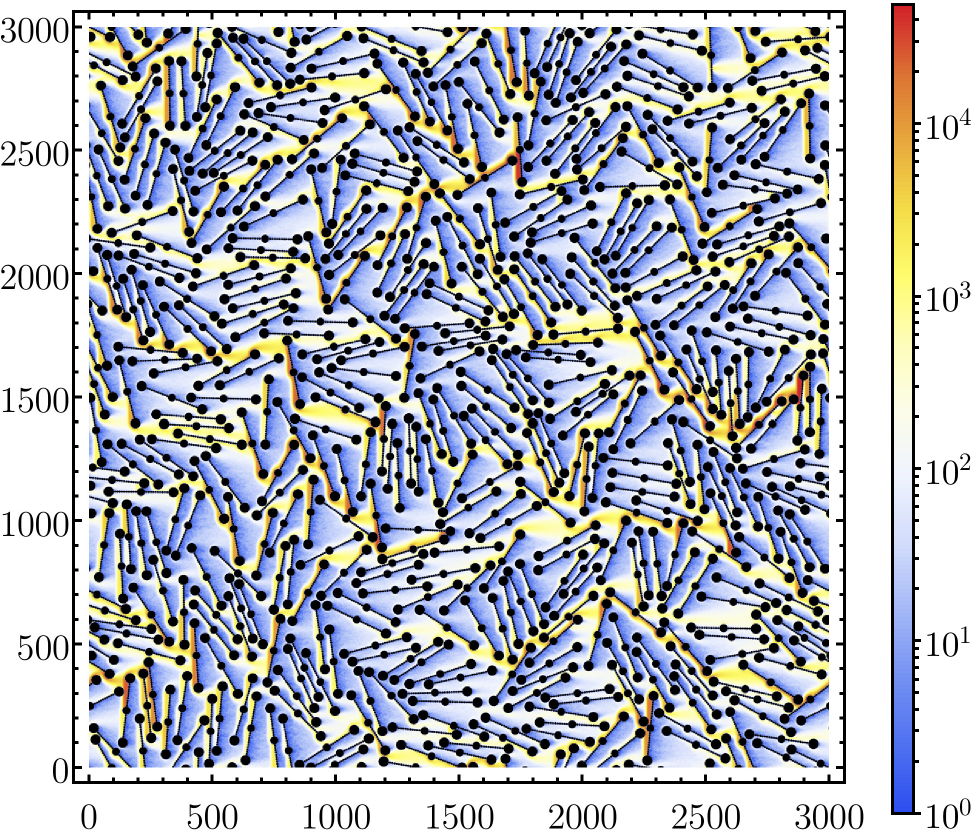}
    \caption{Frequency count of distinct visits at $(x,y)$ points for $5 \times 10^5$ trajectories of the duration $T=10^5$.
    Simulation parameters: deterministic drift $\Lambda=1$, standard deviation of the orientational angle $\sigma(\theta)=0^\circ$ (top panel), $\sigma(\theta)=20^\circ$ (middle panel) and $\sigma(\theta)=180^\circ$ (bottom panel).
    }
    \label{fig:2dvisited}
\end{figure}
\begin{figure}[!h]
    \centering
    \includegraphics[width=\columnwidth]{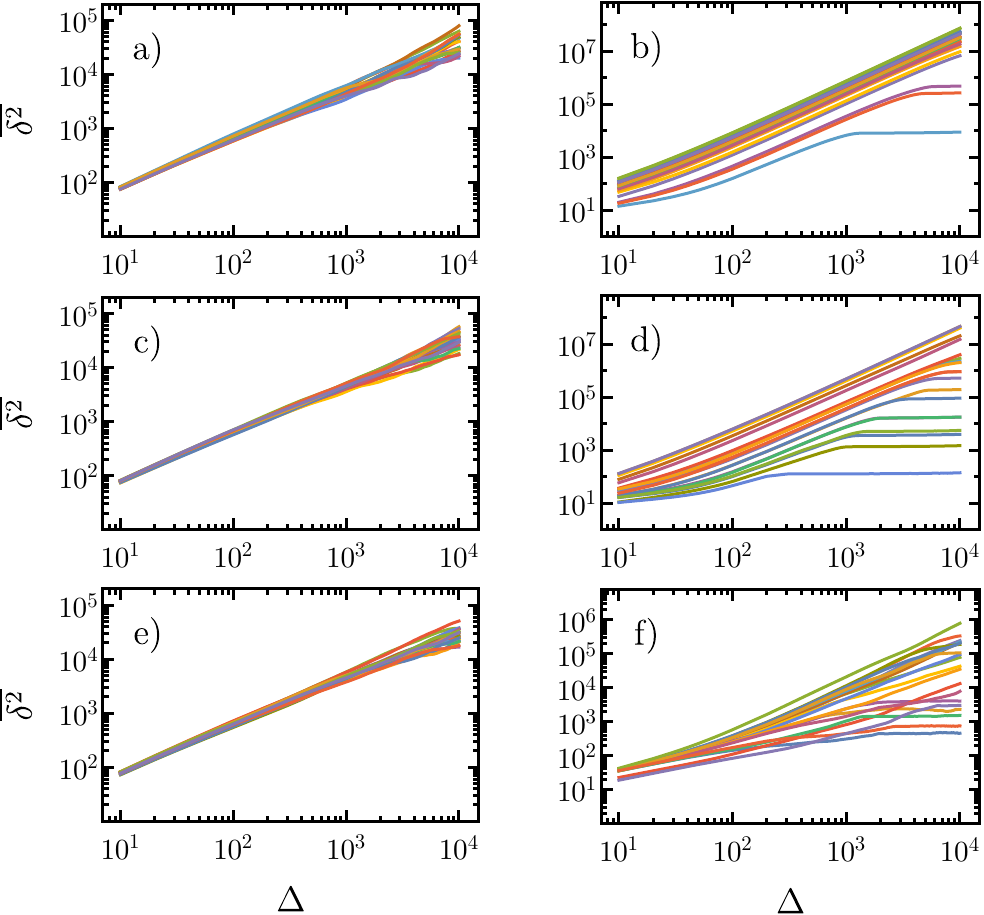} \\
    \caption{Sample individual time-averaged mean square displacements $\overline{\delta^2(\Delta,T)}$ in the absence ($\Lambda=0$, left column) and presence ($\Lambda=1$, right column) of the drift.
    Various rows correspond to various values of the standard deviation $\sigma(\theta)$ of the orientational angle $\theta$: $\sigma(\theta)\in\{0^\circ,20^\circ,180^\circ\}$  (from top to bottom).
    }
    \label{fig:tamsd}
\end{figure}

The heterogeneity of the environment is responsible for the inhomogeneity of diffusive patterns, which is already visible on the single trajectory level as various paths differ very much, see Fig.~\ref{fig:environment}.
The inhomogeneity of the environment can be quantified by  counting distinct visited points.
Fig.~\ref{fig:2dvisited} shows the number of distinct visits to each point $(x,y)$ for $\Lambda=1$ and various level of ordering, corresponding to different values of the standard deviation $\sigma(\theta)$, $\sigma(\theta) \in \{0^\circ,20^\circ,180^\circ\}$ (from top to bottom).
For $\sigma(\theta)=180^\circ$ the standard deviation of the orientation angle $\theta$ of fibrinogen molecules is maximal.
Nevertheless, due to properties of the RSA method, some level of alignment is still visible.
The partial ordering is produced by the excluded volume effect, more precisely during building of the environment, randomly placed particles cannot overlap.
Consequently, some neighboring fibrinogens are close to being parallel.
To reduce the role of traps, for each trajectory multiple visits to the same point are uncounted, i.e., only distinct visits from various trajectories are counted.
More precisely, for every trajectory, if a point $(x,y)$ is visited more than once, the first visit is counted only.
Otherwise, regions in which a particle can be trapped will dominate over other points.
Histograms depicted in Fig.~\ref{fig:2dvisited} have been estimated using $5\times 10^5$ trajectories of the length $10^5$.
They prove that the action of the drift produces pathways because for $\Lambda=0$ densities of distinct visited points are uniform (results not shown).
%
Fig.~\ref{fig:2dvisited} also demonstrates that some restricted parts of the domain of motion are well filled with distinct visited points by diffusing particles.
This is especially well visible for the highest disorder, i.e., in the bottom panel of Fig.~\ref{fig:2dvisited} which corresponds to $\sigma(\theta)=180^\circ$.
Importantly, the presence of fibrinogen makes some parts of the space inaccessible.
A particle can get trapped, making the individual trajectories very different and sensitive to initial conditions, which is demonstrated by the the irregularity of pathways visible in Fig.~\ref{fig:2dvisited} and irregularity of trajectories depicted in Fig.~\ref{fig:environment}.
%
%
The main source of this sensitivity is in the inhomogeneity of the environment. In order to analyze this phenomenon in more detail, we have inspected individual trajectories. Fig.~\ref{fig:tamsd} presents sample realizations of time-averaged mean square displacement, see Eq.~(\ref{eq:tamsd}), in the absence ($\Lambda=0$, left column) and presence ($\Lambda=1$, right column) of drift for various orderings of the environment.
Subsequent rows of Fig.~\ref{fig:tamsd} contain results corresponding to various values of the standard deviation $\sigma(\theta)$ of the orientational angle $\theta$ between fibrinogen and $x$-axis: $\sigma(\theta) \in \{0^\circ,20^\circ,180^\circ\}$ (from top to bottom).
Due to presence of obstacles, the individual time-averaged mean square displacements significantly vary between realizations.
The addition of drift substantially increases the level of variability, compare the left ($\Lambda=0$) and right ($\Lambda=1$) columns of Fig.~\ref{fig:tamsd}.
Under the action of drift chances of trapping the particle are significantly larger than in the absence of drift, i.e., a significant fraction of time-averaged mean square displacements saturates.
Moreover, increased chances of trapping are reflected in the increased level of variability among individual time-averaged mean square displacements $\overline{\delta^2(\Delta,T)}$.

\begin{figure}[!h]
    \centering
    \includegraphics[width=\columnwidth]{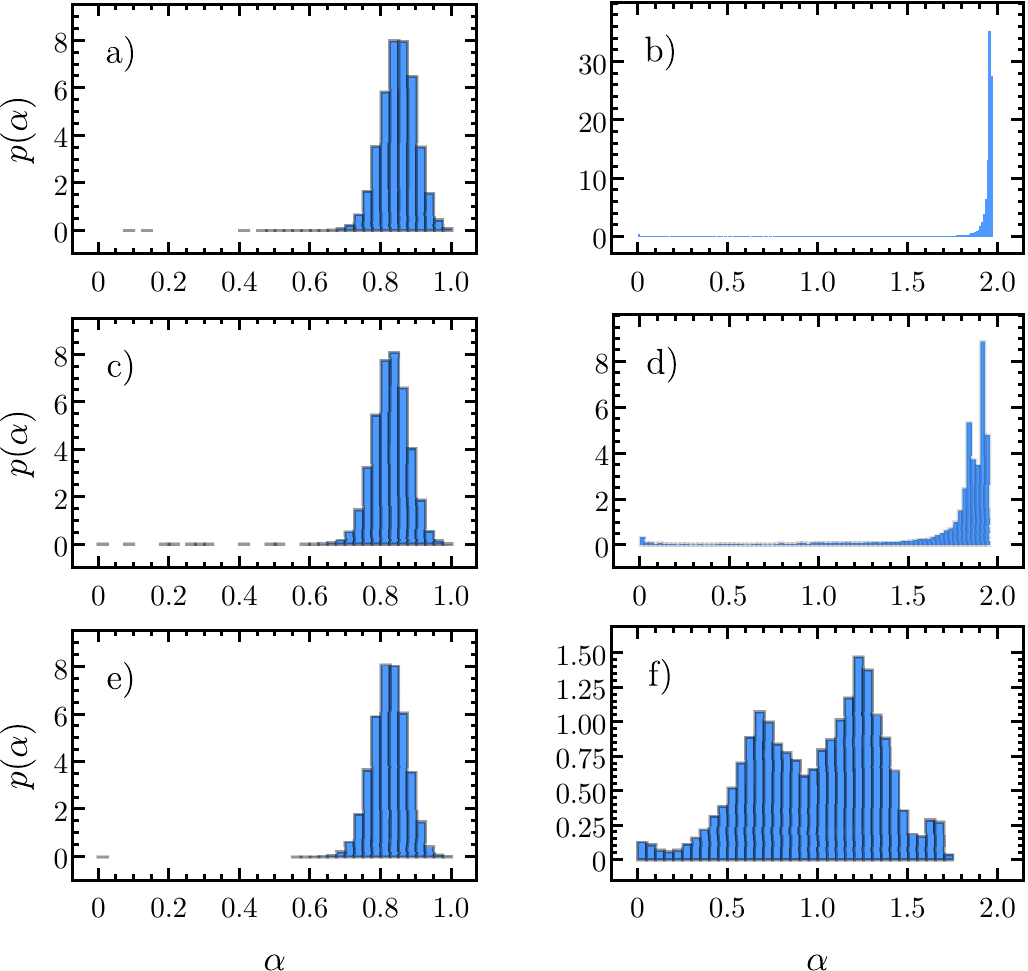} \\
    \caption{Histograms of exponents $\alpha$ fitted to individual time-averaged mean square displacements, see Eq.~(\ref{eq:tamsdfit}) and Fig.~\ref{fig:tamsd}, in the absence ($\Lambda=0$,  left column) and presence ($\Lambda=1$, right column) of the drift, computed for $10^5$ trajectories.
    Various rows correspond to various values of $\sigma(\theta)$: $\sigma(\theta)\in\{0^\circ,20^\circ,180^\circ\}$  (from top to bottom).
    }
    \label{fig:tamsdalphahist}
\end{figure}

The level of variability among individual trajectories can be further explored by inspection of exponents characterizing various individual trajectories. From the individual time-averaged mean square displacement, it is possible to estimate histograms of these exponents
\begin{equation}
 \overline{\delta^2(\Delta,T)} \propto \Delta^\alpha.
 \label{eq:tamsdfit}
\end{equation}
The exponent $\alpha$ is estimated by fitting the power-law dependence to $\overline{\delta^2(\Delta,T)}$ for $\Delta \in [10^2, 10^3]$, see Fig.~\ref{fig:tamsd}. The choice of the specific range is motivated by the study of $\overline{\delta^2(\Delta,T)}$ for the Brownian diffusion in the empty space.
In such a case, for $\Delta > 10^3$, due to the finite observation time $T$, time-averaged mean square displacements start to vary significantly between trajectories.
Numerically calculated histograms are depicted in Fig.~\ref{fig:tamsdalphahist}.
In the absence of drift, as measured by the scaling of the time-averaged mean square displacement, the process is subdiffusive, i.e., the recorded values of $\alpha$ are smaller than 1, see left column of Fig.~\ref{fig:tamsdalphahist}.
Moreover, with the increasing standard deviation $\sigma(\theta)$ of the orientational angle $\theta$ exponents $\alpha$ slightly decrease, see top left and bottom left panels of Fig.~\ref{fig:tamsdalphahist} which prove that the modal value decreases with the increasing $\sigma(\theta)$.
Action of the deterministic constant force (drift) results in observations of larger values of exponents $\alpha$ characterizing the growth of the time-averaged mean square displacements, see right column of Fig.~\ref{fig:tamsdalphahist}.
With the increasing standard deviation $\sigma(\theta)$ of the orientational angle $\theta$ between the fibrinogen and $x$-axis, exponents $\alpha$ become more dispersed as individual trajectories $(x(t),y(t))$ and individual time-averaged mean square displacements $\overline{\delta^2(\Delta,T)}$ are more distinctive.
Moreover, $\alpha$ smaller than two are more likely to be observed, because obstacles prevent long-lasting ballistic motion.
Finally, for $\sigma(\theta)=180^\circ$, the histogram of observed exponents $\alpha$ is multimodal and very broad because of the high level of heterogeneity.
Naive interpretation of recorded exponents might suggest that the motion can be superdiffusive, as $\alpha$  can be larger than 1.
The actual reason why the exponents $\alpha>1$ are observed is the competition between ballistic motion and restrictions on possible displacements introduced by obstacles.
Nevertheless, as it is already visible in Figs.~\ref{fig:2dhis} and~\ref{fig:tamsd}, the action of the drift in the environment full of obstacles is quite different than in the empty space.

\begin{figure}[!h]
    \centering
    \includegraphics[width=\columnwidth]{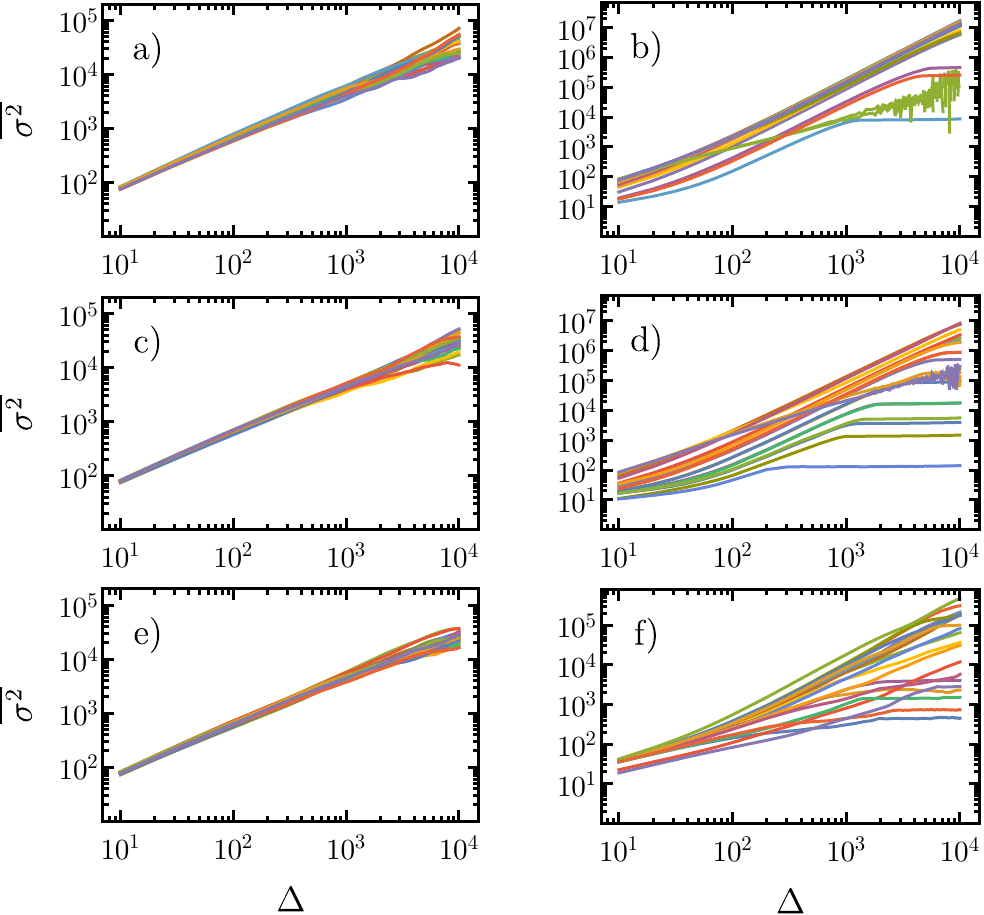} \\
    \caption{Sample individual time-averaged variances  $\overline{\sigma^2(\Delta,T)}$, see Eq.~(\ref{eq:tavariance}), in the absence ($\Lambda=0$, left column) and presence ($\Lambda=1$, right column) of the drift.
    Various rows correspond to various values of the standard deviation $\sigma(\theta)$ of the orientational angle $\theta$: $\sigma(\theta)\in\{0^\circ,20^\circ,180^\circ\}$  (from top to bottom).}
    \label{fig:tavar}
\end{figure}

\begin{figure}[!h]
    \centering
    \includegraphics[width=\columnwidth]{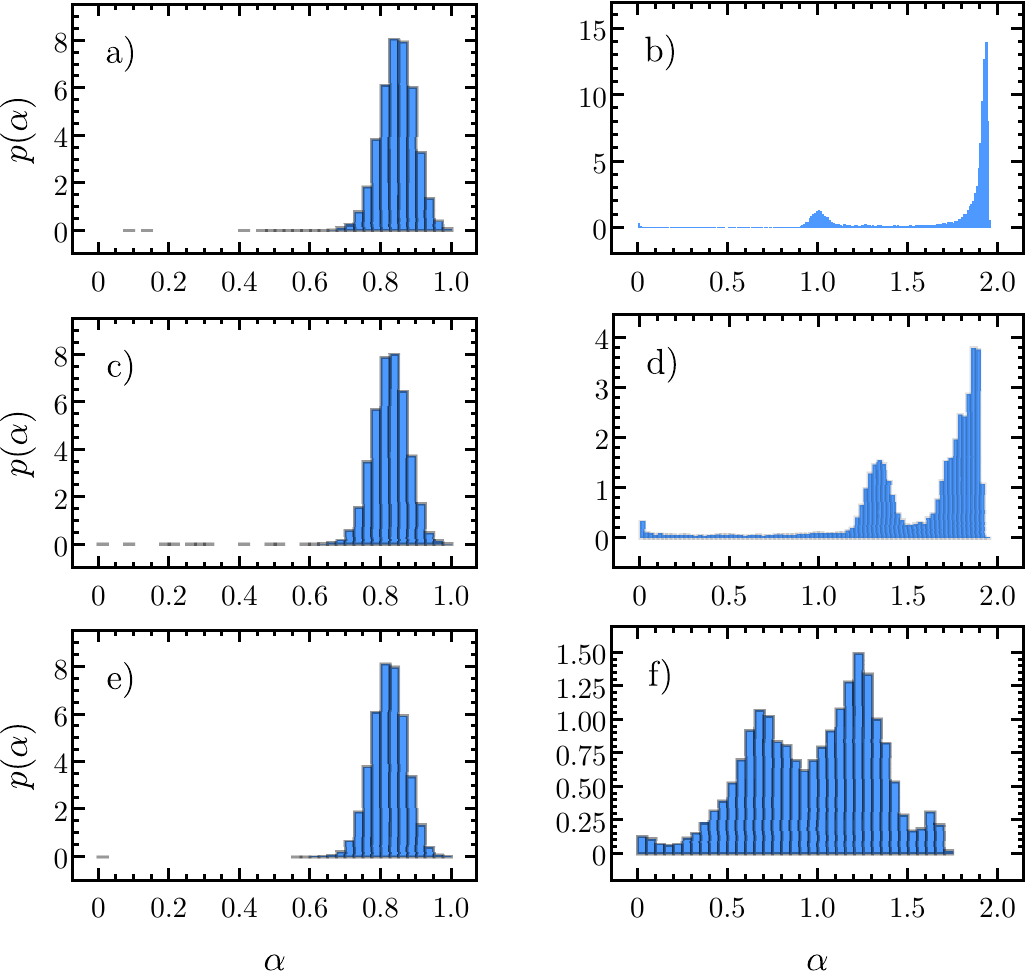} \\
     \caption{Histograms of exponents $\alpha$ fitted to individual time-averaged variances, see Eq.~(\ref{eq:tavariance}) and Fig.~\ref{fig:tavar}, in the absence ($\Lambda=0$,  left column) and presence ($\Lambda=1$, right column) of the drift, computed for $10^5$ trajectories.
    Various rows correspond to various values of $\sigma(\theta)$: $\sigma(\theta)\in\{0^\circ,20^\circ,180^\circ\}$  (from top to bottom).
    }
    \label{fig:tavaralphahist}
\end{figure}

In the empty space, the ballistic motion is produced by the drift.
In such a case, under Gaussian white noise driving, the mean square displacement for detrended trajectories reveals the normal diffusion.
Therefore, the apparent superdiffusion can be discriminated by the subtraction of the average trend.
In the empty space, the analysis of detrended trajectories is equivalent to the analysis of variance (1D, see Eq.~(\ref{eq:std1d})) or trace of the covariance matrix (2D, see Eq.~(\ref{eq:std2d})).
In the crowded environment, the situation becomes more complex, as the presence of obstacles terminates the deterministic motion  produced by the constant drift and makes individual trajectories very different.
Furthermore, Eqs.~(\ref{eq:std1d}) and~(\ref{eq:std2d}) are strongly associated with the ensemble averaging.
Therefore, in analogy to the time-averaged mean square displacement, see Eq.~(\ref{eq:tamd}), it is necessary to develop a way of detrending a single trajectory.
From the time-averaged mean square displacement, see Eq.~(\ref{eq:tamsd-app}), and time-averaged mean displacement, see Eq.~(\ref{eq:tamd}), it is possible to calculate the analog of time-averaged variance, see Eq.~(\ref{eq:tavariance}).
Examination of the time-averaged variance for a free particle in the absence of drift or under the action of constant drift correctly discriminates the type of diffusive part of the motion.
Moreover, in the driftless case, the analysis of time-averaged variance and time-averaged mean square displacement in crowded environments, gives the same results, see
left columns of Figs.~\ref{fig:tamsd} and~\ref{fig:tavar} showing time-averaged dependencies and left columns of Figs.~\ref{fig:tamsdalphahist} and~\ref{fig:tavaralphahist} depicting histograms of fitted exponents $\alpha$.
Comparison of left panels of Figs.~\ref{fig:tamsd} and~\ref{fig:tavar} as well as Figs.~\ref{fig:tamsdalphahist} and~\ref{fig:tavaralphahist}  shows that the driftless case can be analyzed by inspection of time-averaged mean square displacement or time-averaged variance because both measures provide consistent results.
The very different situation is observed under the constant drift.
This time, the individual curves in the right panels of Figs.~\ref{fig:tamsd} and~\ref{fig:tavar} are not so different between the two figures.
Nevertheless, the differences are very well visible in the histograms of obtained exponents $\alpha$, see right columns of Figs.~\ref{fig:tamsdalphahist} and~\ref{fig:tavaralphahist}.
Both histograms show values of the exponent $\alpha$ indicating superdiffusion.
The overall trend cannot be successfully subtracted, because the presence of obstacles makes deterministic trends transient.
After a sufficiently long time, the particle hits the fibrinogen and the deterministic motion becomes hindered.
Moreover, it makes the whole process of detrending problematic, because it subtracts the average trend also from trapped particles inducing spurious movement, what is manifested by the multimodality of $p(\alpha)$ densities in the top and middle right panels of Fig.~\ref{fig:tavaralphahist}.
With the increasing strength of the drift, the duration of the deterministic motion decreases.
Therefore, spread of exponents $\alpha$ is produced by the combined action of drift and obstacles.

\begin{figure}[!h]
    \centering
    \includegraphics[width=0.9\columnwidth]{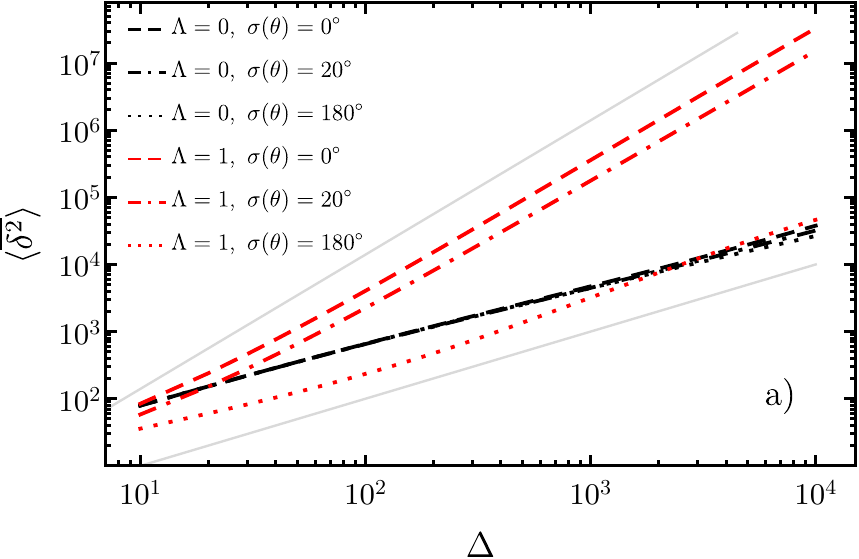} \\
    \includegraphics[width=0.9\columnwidth]{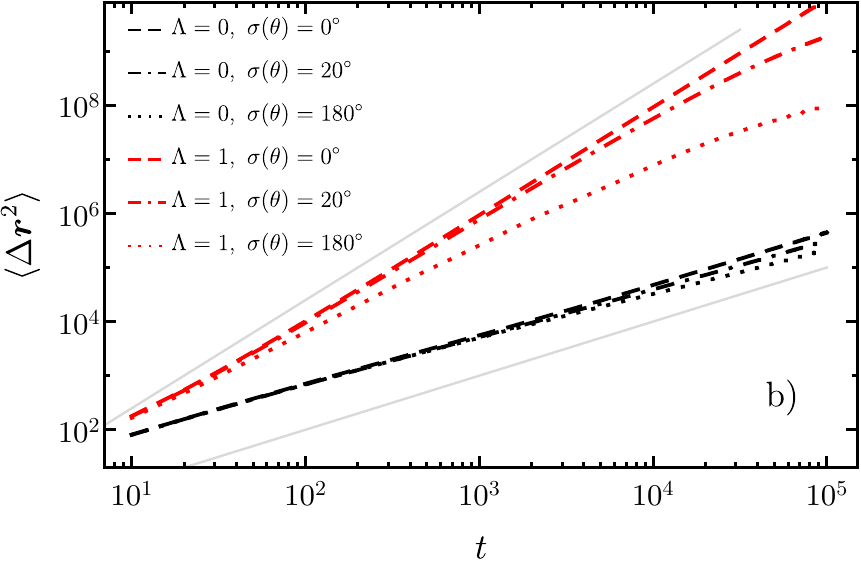} \\
    \caption{
    Ensemble averaged time-averaged mean square displacement, i.e., $\langle  \overline{\delta^2(\Delta,T)} \rangle$, (top panel --- a) and the mean square displacement, i.e., $\langle \Delta\bm{r}^2(t) \rangle$ (bottom panel --- b)
    in the absence [black (darker)] and presence [red (lighter)] of the drift. Dashed, dot-dashed and dotted lines correspond to respective values of $\sigma(\theta)$: $\sigma(\theta)\in\{0^\circ,20^\circ,180^\circ\}$. Light gray lines depict linear and quadratic functions.
    }
    \label{fig:eatmsd}
\end{figure}

Figure~\ref{fig:eatmsd} shows the ensemble averaged time-averaged mean square displacement calculated from the ensemble of observations \cite{metzler2014anomalous}, i.e.,
$\langle  \overline{\delta^2(\Delta,T)} \rangle$, see Eq.~(\ref{eq:tamsd}).
Exponents $\alpha$ fitted to individual curves are presented in the left part of Tab.~\ref{tab:alpha}.
For $\Lambda=0$ the obtained values of exponents are smaller than 1, while for $\Lambda=1$ they are larger than 1.
The (Markovian) subdiffusion observed for $\Lambda=0$ is due to the fractal character of the environment \cite{bouchaud1990,saxton1994anomalous,ciesla2015taming}, while the apparent superdiffusion for $\Lambda=1$ is due to obstacle-terminated  ballistic motion.
With increasing $\sigma(\theta)$, the disorder of the environment grows, slowing the diffusion.
This in turn is responsible for the decrease in the exponent $\alpha$, see left part of Tab.~\ref{tab:alpha}.
%

\begin{table}[!h]
    \centering
    \begin{tabular}{c||c|c||c|c}
    & \multicolumn{2}{c||}{$\langle \overline{\delta^2} \rangle$} & \multicolumn{2}{c}{$\langle \Delta \bm{r}^2 \rangle$}  \\
    $\sigma(\theta)$ & $\;\Lambda=0\;$ &  $\Lambda=1$ & $\;\Lambda=0\;$ &  $\Lambda=1$ \\ \hline\hline
    $0^\circ$ & 0.92 & 1.95 & 0.91 & 1.88\\ \hline
    $20^\circ$ & 0.86   & 1.94 & 0.84 & 1.67 \\ \hline
    $180^\circ$ & 0.76 & 1.10 & 0.76 & 0.61\\
    \end{tabular}
    \caption{Values of fitted exponents $\alpha$ to the ensemble averaged time-averaged mean square displacement $\langle \overline{\delta^2(\Delta,T)} \rangle$, see Fig.~\ref{fig:eatmsd}a, and mean square displacement $\langle \Delta \bm{r}^2(t) \rangle $, see Fig.~\ref{fig:eatmsd}b.}
    \label{tab:alpha}
\end{table}

The ensemble averaged time-averaged mean square displacement $\langle \overline{\delta^2(\Delta,T)} \rangle$, see Fig.~\ref{fig:eatmsd}a and left part of Tab.~\ref{tab:alpha}, can be contrasted with the mean square displacement $\langle \Delta\bm{r}^2(t) \rangle$, see Fig.~\ref{fig:eatmsd}b and right part of Tab.~\ref{tab:alpha}.
Comparison of the ensemble averaged time-averaged mean square displacement and the mean square displacement, cf., Fig.~\ref{fig:eatmsd}a and Fig.~\ref{fig:eatmsd}b, proves that  in the crowded environment the weak ergodicity breaking is observed.
Action of the constant drift further increases the level of weak ergodicity breaking.
This effect is further confirmed by distinct values of the fitted exponents, see  Tab.~\ref{tab:alpha}.

Ensemble averaged time-averaged mean square displacement, see Fig.~\ref{fig:eatmsd}a, and the mean square displacement, see Fig.~\ref{fig:eatmsd}b, shows very different patterns.
The trapping of particles with $\Lambda>0$ is very well visible in Fig.~\ref{fig:eatmsd}b where the mean square displacements $\langle \Delta \bm{r}(t) \rangle$ after some time (dependent on the $\sigma(\theta)$) start to bend.
At the same time, the ensemble averaged time-averaged mean square displacements, in log-log scale, follow almost straight lines, see Fig.~\ref{fig:eatmsd}a.
The difference in the shape of MSD and ensemble averaged TAMSD originates in a delayed response of TAMSD to trapping, see Eq.~(\ref{eq:tamsd}).
This can be illustrated by studying the artificial, piecewise linear and constant, e.g., $x(t)=\min\{t,10\}$  trajectory.
For such a trajectory, the mean square displacement immediately saturates at $t=10$, while time-averaged mean square displacement still grows.

\begin{figure}[!h]
    \centering
    \includegraphics[width=0.9\columnwidth]{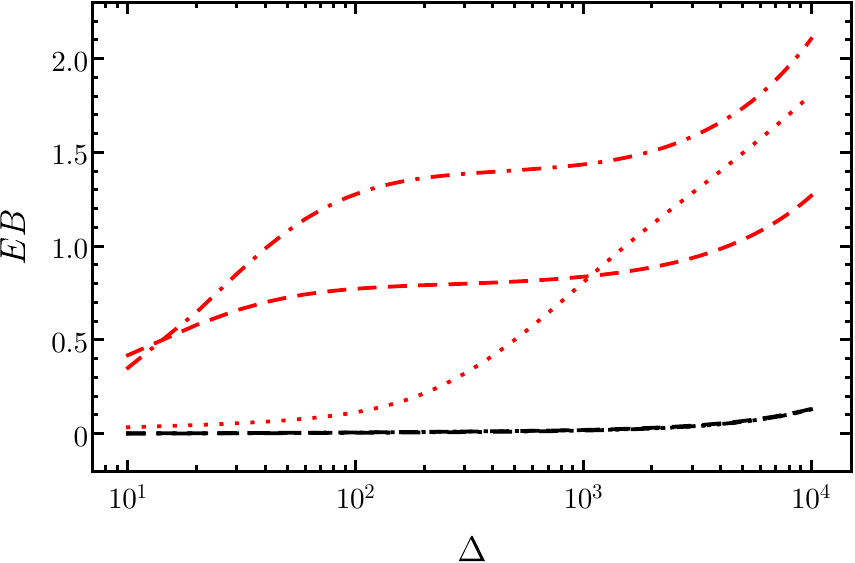} \\
    \caption{
    Time dependence of the ergodicity breaking parameter, see Eq.~(\ref{eq:eb}), in the absence [black (darker)] and presence [red (lighter)] of the drift, see Fig.~\ref{fig:eatmsd}. Dashed, dot-dashed and dotted lines correspond to respective values of $\sigma(\theta)$: $\sigma(\theta)\in\{0^\circ,20^\circ,180^\circ\}$).
    Please note that in the absence of drift ($\Lambda=0$) all  [black (darker)] curves coincide.
        }
    \label{fig:eb}
\end{figure}

Finally, the ergodicity breaking can be quantified by the ergodicity breaking (EB) parameter \cite{he2008,burov2011single}
\begin{equation}
EB =     \lim_{T\to\infty} \frac{\left\langle \left(\overline{\delta^2}\right)^2 \right\rangle -\left\langle \overline{\delta^2} \right\rangle^2}{ \left\langle \overline{\delta^2} \right\rangle ^2 },
\label{eq:eb}
\end{equation}
which is depicted in Fig.~\ref{fig:eb}.
Fig.~\ref{fig:eb} clearly shows that the action of the drift significantly increases the level of ergodicity breaking in comparison to the driftless case.
Moreover, the level of ergodicity breaking grows in time.
For $\Lambda>0$, the growth rate is significantly larger than for the driftless case.

\bdt{Finally, in order to verify the role played by the concave parts of the fibrinogen molecules, see Fig.~\ref{fig:environment}, we have repeated all simulations for spherocylinders of the same length and area as fibrinogen molecules.
From these simulations, it is visible that the conclusions drawn for the crowded environment built by fibrinogen particles are quantitatively unaffected.
The lack of concave part of obstacles prevents building of traps for the ordered setups and stops the accumulation of particles in cavities.
Consequently, the diffusion is slightly accelerated and the widths of histograms of observed exponents $\alpha$ are decreased.
Importantly, these additional simulations prove that the drift is the main factor responsible for the plenitude of observed effects.
Representative figures corresponding to Figs.~\ref{fig:2dvisited},
\ref{fig:tamsd},
\ref{fig:tamsdalphahist},
\ref{fig:eb}
are included in Appendix~\ref{sec:spherocylinders}.
}




\section{Summary\label{sec:summary}}

Diffusion is one of the most widespread processes, which typically is driven by the gradient of concentration.
It underlines the plenitude of everyday phenomena, like equilibration of concentration, the motion of cultural traits across geographical areas, isotope separation, or money spread, to name a few.

From the microscopic point of view, diffusion originates due to constant interactions among particles.
Therefore, the theory of stochastic processes provides a natural framework for the description of diffusive processes.
On the microscopic level, diffusion can be described by the Langevin equation, while on the macroscopic level, by the diffusion equation.
The environment can significantly affect properties of diffusion.
In particular, the diffusion in the crowded environment can display a signature of subdiffusion because of fractal structure of the environment.
Due to space inhomogeneity, individual trajectories display very large variability.
Consequently, the signatures of diffusion are  sensitive to initial conditions and manifest differences between trajectories and the system break ergodicity in the weak sense.

Addition of drift further changes characteristics of diffusion.
In the empty space, the constant drift is responsible for the emergence of the deterministic trend, i.e., the ballistic motion.
In the crowded environment, the deterministic trend is very transient because presence of obstacles efficiently prevents unlimited deterministic motion.
The action of drift significantly enlarges the chances of trapping and increases the level of variability among trajectories.
\bdt{As measured by the properties of individual time-averaged mean square displacements, the action of the drift can not only induce superdiffusive scaling, but it can also enhance the subdiffusivity.
Enhanced subdiffusivity is manifested by the observation of smaller values of the exponent characterizing the growth of individual time-averaged mean square displacements than in driftless setups.}
Furthermore, the action of the drift amplifies the level of weak ergodicity breaking.

%
%
\appendix
\section{Langevin equation\label{sec:langevin}}
In order to elucidate the main properties of diffusive systems, we recall basic information about the motion of a free particle under the action of Gaussian white noise (GWN).
These information are provided to underline differences between a free motion and a motion with a drift.
The motion of a free particle under the action of GWN is described by the following Langevin equation
\begin{equation}
    \frac{dx(t)}{dt}=\xi(t),
    \label{eq:langevin}
\end{equation}
where $\xi(t)$ is the Gaussian white noise satisfying
\begin{equation}
    \langle \xi(t) \rangle=0 \;\;\;\;\; \mbox{and} \;\;\;\;\;    \langle \xi(t)\xi(s) \rangle=\sigma_0^2 \delta(t-s).
    \label{eq:gwn}
\end{equation}
The Langevin equation~(\ref{eq:langevin}) defines trajectories of the Wiener process (Brownian motion) \cite{gardiner2009}.
Eq.~(\ref{eq:langevin}) is associated with the Smoluchowski--Fokker--Planck equation \cite{risken1996fokker,gardiner2009}, which describes the evolution of the probability density $p=p(x,t|x_0,t_0)=\langle \delta(x-x(t)) \rangle $
\begin{equation}
    \frac{\partial p}{\partial t}= \frac{\sigma_0^2}{2} \frac{\partial^2 p}{\partial x^2}
    \label{eq:sfp}.
\end{equation}
Under the initial condition $x(0)=0$, Eq.~(\ref{eq:sfp}) has the solution given by the normal distribution with the zero mean and the linearly increasing variance
\begin{equation}
\sigma^2(t)=\sigma_0^2 t,
\end{equation}
i.e.,
\begin{equation}
    p(x,t)=\frac{1}{\sqrt{2\pi \sigma^2(t)}} \exp \left[ -\frac{x^2}{2\sigma^2(t)}  \right].
    \label{eq:sfp-solution}
\end{equation}

From the formal solution of Eq.~(\ref{eq:langevin})
\begin{equation}
    x(t)=\int_0^t \xi(s)ds
\end{equation}
and~Eq.~(\ref{eq:gwn}) or from Eq.~(\ref{eq:sfp-solution}) one can calculate the mean square displacement $\langle \Delta x^2 (t) \rangle$
\begin{eqnarray}
    \langle \Delta x^2 (t) \rangle & = &  \langle [x(t)-x(0)]^2 \rangle \\ \nonumber
    & = &  \int [x(t)-x(0)]^2 p(x,t) dx= \sigma_0^2 t \propto t.
\end{eqnarray}
The mean square displacement (MSD) is used as the standard method of discriminating the diffusion type \cite{metzler2000,metzler2004}.
The linear scaling of the MSD along with the definition of $x(t)$ confirms that the process $x(t)$, see Eq.~(\ref{eq:langevin}), exhibits normal Markovian diffusion.

The situation gets more complex in the case of a constant drift $\Lambda$.
The Langevin equation
\begin{equation}
    \frac{dx(t)}{dt}=\Lambda+\xi(t),
    \label{eq:langevin-drift}
\end{equation}
is associated with the Smoluchowski--Fokker--Planck \cite{risken1996fokker,gardiner2009} equation $p=p(x,t|x_0,t_0)$
\begin{equation}
    \frac{\partial p}{\partial t}= -\frac{\partial}{ \partial x} \left[ \Lambda p \right] + \frac{\sigma_0^2}{2} \frac{\partial^2 p}{\partial x^2}
    \label{eq:sfp-drift},
\end{equation}
which, for $x(0)=0$, has the solution given by the following normal density
\begin{equation}
    p(x,t)=\frac{1}{\sqrt{2\pi \sigma^2(t)}} \exp \left[ -\frac{\left(x-\mu(t)\right)^2}{2\sigma^2(t)}  \right],
    \label{eq:sfp-drift-solution}
\end{equation}
where
\begin{equation}
\sigma^2(t)=\sigma_0^2 t
\end{equation}
and
\begin{equation}
\mu(t)=\Lambda t
\end{equation}
are the variance and the mean, respectively.
For such process, the MSD is given by
\begin{eqnarray}
\label{eq:msd1d}
    \langle \Delta x^2 (t) \rangle & = &  \langle [x(t)-x(0)]^2 \rangle \\ \nonumber
     & = &  \int  [x(t) -x(0)]^2 p(x,t) dx \\ \nonumber
     &=&\Lambda^2 t^2 + \sigma_0^2 t.
\end{eqnarray}
At the same time
\begin{eqnarray}
\label{eq:std1d}
    \mathrm{var}(x(t)) & = & \sigma^2(t) = \langle [x(t)-\langle x(t) \rangle]^2 \rangle \\ \nonumber
     & = &  \langle [x(t)- \Lambda t]^2 \rangle = \sigma_0^2 t.
\end{eqnarray}

Equation~(\ref{eq:langevin-drift}) can be easily generalized to the 2D setup
\begin{equation}
    \frac{d \bm{r}(t)}{dt}=\bm{\Lambda}+\bm{\xi}(t),
    \label{eq:langevin2d-app}
\end{equation}
where $\bm{\Lambda}$ represents the constant drift and $\bm{\xi}(t)$ is a 2D Gaussian white noise \cite{samorodnitsky1994}.
For a spherically symmetric 2D GWN
\begin{equation}
\bm{\xi}(t)=(\xi_x(t),\xi_y(t)),
\end{equation}
where $\xi_x(t)$ and $\xi_y(t)$ are two independent 1D Gaussian white noises satisfying Eq.~(\ref{eq:gwn}) and the additional condition
\begin{equation}
    \langle \xi_x(t)\xi_y(s) \rangle=0.
\end{equation}
Both conditions can be written in the compact form
\begin{equation}
    \langle \xi_i(t) \xi_j(s) \rangle = \sigma_0^2 \delta_{ij} \delta(t-s),
\end{equation}
where $\{i,j\}\in\{x,y\}$ and $\delta_{ij}$ is the Kronecker delta.
The Langevin equation~(\ref{eq:langevin2d-app}) is associated with the Smoluchowski--Fokker--Planck equation \cite{risken1996fokker,gardiner2009}
\begin{equation}
    \frac{\partial p}{\partial t}= -\bm{\Lambda} \cdot \left[ \nabla  p \right] + \frac{\sigma_0^2}{2} \nabla^2 p
    \label{eq:sfp-drift-2d}.
\end{equation}
The diffusion equation~(\ref{eq:sfp-drift-2d}) has the following solution
\begin{eqnarray}
p(\bm{r},t) & = & \frac{1}{2\pi \sigma^2(t)} \exp \left[ -\frac{ (\bm{r} -\bm{\Lambda})^2 }{2\sigma^2(t)}  \right] \\ \nonumber
    & = & \frac{1}{2\pi \sigma^2(t)} \exp \left[ -\frac{ (x -\Lambda_x)^2 + (y -\Lambda_y)^2}{2\sigma^2(t)}  \right],
    \label{eq:sfp-drift-2d-solution}
\end{eqnarray}
where, as previously, $\sigma^2(t)=\sigma_0^2 t$.
For the model described by Eq.~(\ref{eq:langevin2d-app}) the covariance matrix $\mathbf{\Sigma}$ is diagonal, i.e.,
\begin{equation}
    \mathrm{var}(x)=\mathrm{var}(y)=\sigma_0^2 t
\end{equation}
and
\begin{equation}
    \mathrm{cov}(x,y)=0.
\end{equation}
In the 2D setup, the mean square displacement can be generalized into
\begin{eqnarray}
\label{eq:msd2d}
\langle \Delta \bm{r}^2 (t) \rangle & =  & \langle  [\bm{r}(t) - \bm{r}(0)]^2 \rangle \\ \nonumber
& = &  \langle \Delta x^2(t) \rangle + \langle \Delta y^2(t) \rangle \\ \nonumber
&= & 2\sigma_0^2 t+\bm{\Lambda}^2 t^2.
\end{eqnarray}
It is also possible to calculate the trace of the covariance matrix $\mathbf{\Sigma}$
\begin{equation}
\label{eq:std2d}
\mathrm{Tr} \mathbf{\Sigma} =  \mathrm{var}(x)+\mathrm{var}(y)=2\sigma_0^2 t=\langle [\bm{r} - \langle \bm{r} \rangle]^2 \rangle,
\end{equation}
where $\langle \bm{r} \rangle=\bm{\Lambda} t$.
Without loss of generality, it is possible to align the drift along the $x$-axis.
In the empty space, analogously like the variance in 1D, the trace of the covariance matrix $\mathbf{\Sigma}$ is insensitive to the constant drift $\bm{\Lambda}$.
At the same time, the mean square displacement is dominated by the constant drift. In more general setups, due to interactions with obstacles, the covariance matrix can display different properties than for a free particle, e.g., it does not need to be diagonal.

%
%
\section{Time averaging and ensemble averaging\label{sec:timeaverages}}

From the set of trajectories, the mean square displacement can be calculated as the ensemble average \cite{leijnse2012diffusion}
\begin{equation}
    \langle \Delta \bm{r}^2 (t) \rangle = \int [\bm{r}(t)-\bm{r}(0)]^2 p(\bm{r},t) d^2\bm{r}.
    \label{eq:emsd-app}
\end{equation}
From a single trajectory it is possible to calculate the time-averaged mean square displacement $\overline{ \delta^2(\Delta ,T)  }$
\begin{equation}
    \overline{ \delta^2(\Delta ,T)  } = \frac{1}{T-\Delta} \int_0^{T-\Delta} [ \bm{r}(t+\Delta) - \bm{r}(t)]^2 dt,
    \label{eq:tamsd-app}
\end{equation}
see Refs.~\cite{lubelski2008b,he2008,sokolov2012}.
In ergodic \cite{he2008,burov2011single,cherstvy2013anomalous} systems
\begin{equation}
\lim_{T\to\infty}\overline{ \delta^2(\Delta = t  ,T)  }     = \langle \Delta \bm{r}^2 (t) \rangle,
\end{equation}
i.e., for the infinite observation time, the time average is equal to the ensemble average.

Analogously to the time-averaged mean square displacement $\overline{ \delta^2(\Delta ,T)  }$, see Eq.~(\ref{eq:tamsd-app}), one can calculate the time-averaged mean displacement $\overline{ \bm{\delta}(\Delta ,T)  }$
\begin{equation}
    \overline{ \bm{\delta}(\Delta ,T)  } = \frac{1}{T-\Delta} \int_0^{T-\Delta} [ \bm{r}(t+\Delta) - \bm{r}(t)] dt,
    \label{eq:tamd}
\end{equation}
which is a vector.
From time-averaged mean square displacement, see Eq.~(\ref{eq:tamsd-app}), and time-averaged mean displacement, see Eq.~(\ref{eq:tamd}), it is possible to calculate time-averaged variance
\begin{equation}
    \overline{ \sigma^2(\Delta ,T)  } = \overline{ \delta^2(\Delta ,T)  } -\left[\overline{ \bm{\delta}(\Delta ,T)  }\right]^2,
    \label{eq:tavariance}
\end{equation}
which, contrary to the time-averaged mean square displacement, does not account for the average trend, which is subtracted.

%
%
\section*{Acknowledgments}
This research was supported in part by PLGrid Infrastructure.


%

%

\section{Spherocylinders\label{sec:spherocylinders}}

\begin{figure}[!h]
    \centering
    \includegraphics[width=0.8\columnwidth]{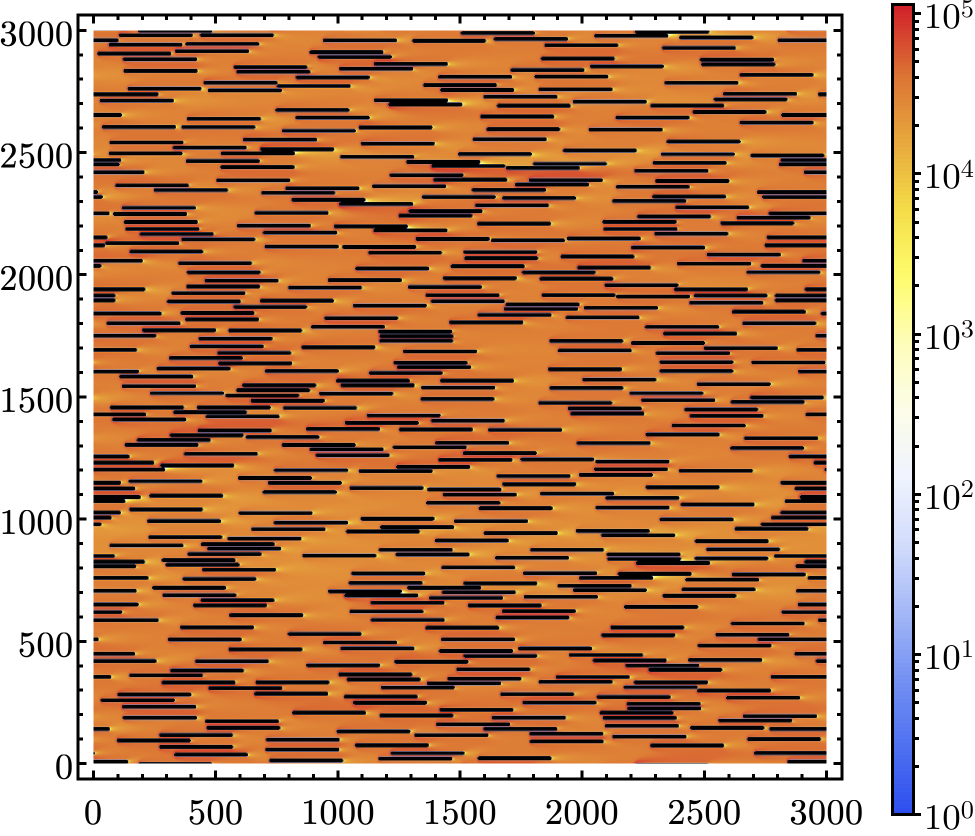} \\
    \vspace{8pt}
    \includegraphics[width=0.8\columnwidth]{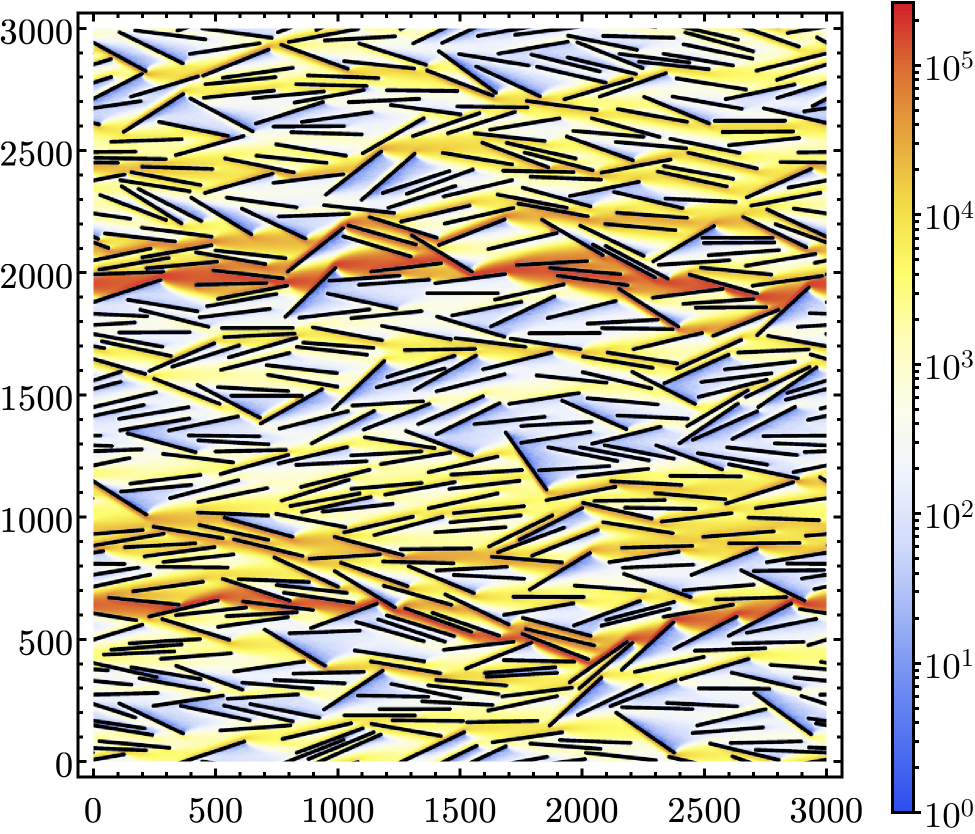}\\
    \vspace{8pt}
    \includegraphics[width=0.8\columnwidth]{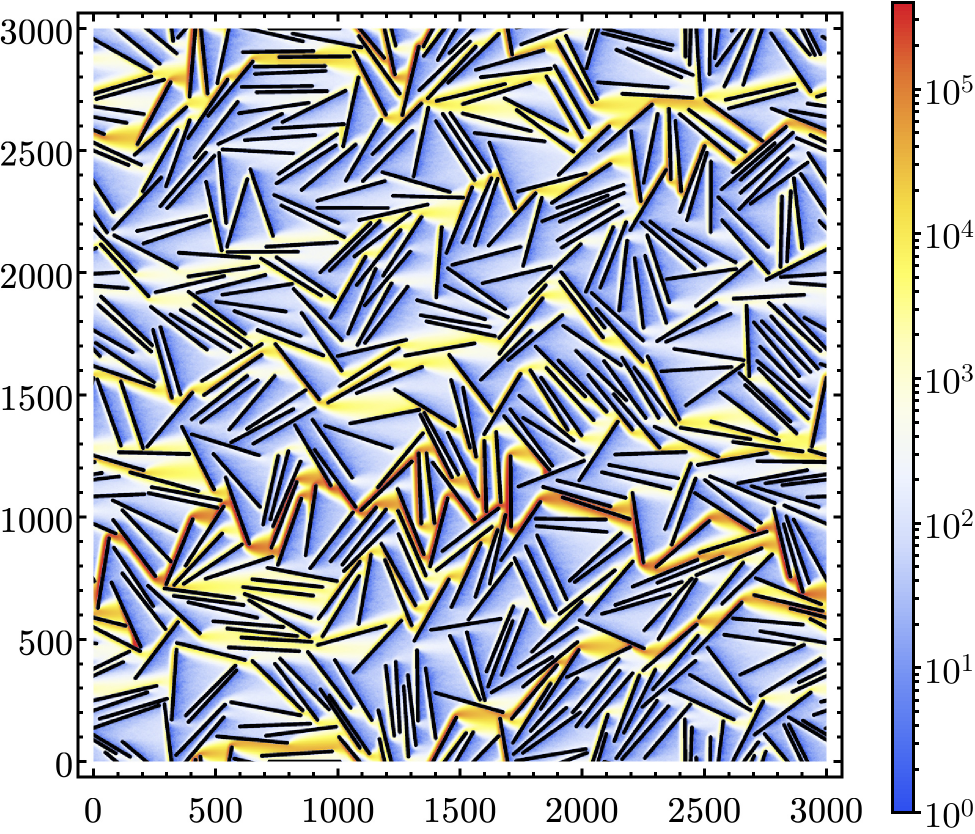}
    \caption{The same as in Fig.~\ref{fig:2dvisited} for spherocylinders.}
    \label{fig:2dvisited-sp}
\end{figure}

\begin{figure}[!h]
    \centering
    \includegraphics[width=\columnwidth]{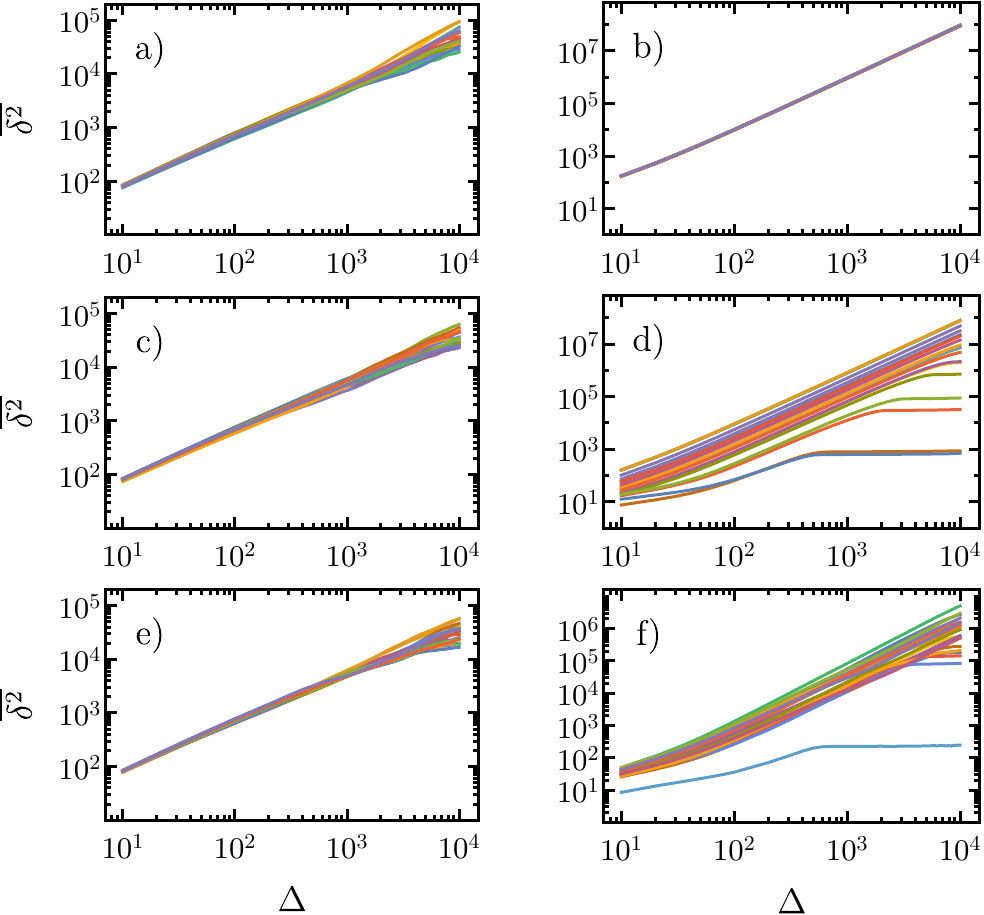} \\
    \caption{The same as in Fig.~\ref{fig:tamsd} for spherocylinders. }
    \label{fig:tamsd-sp}
\end{figure}

\begin{figure}[!h]
    \centering
    \includegraphics[width=\columnwidth]{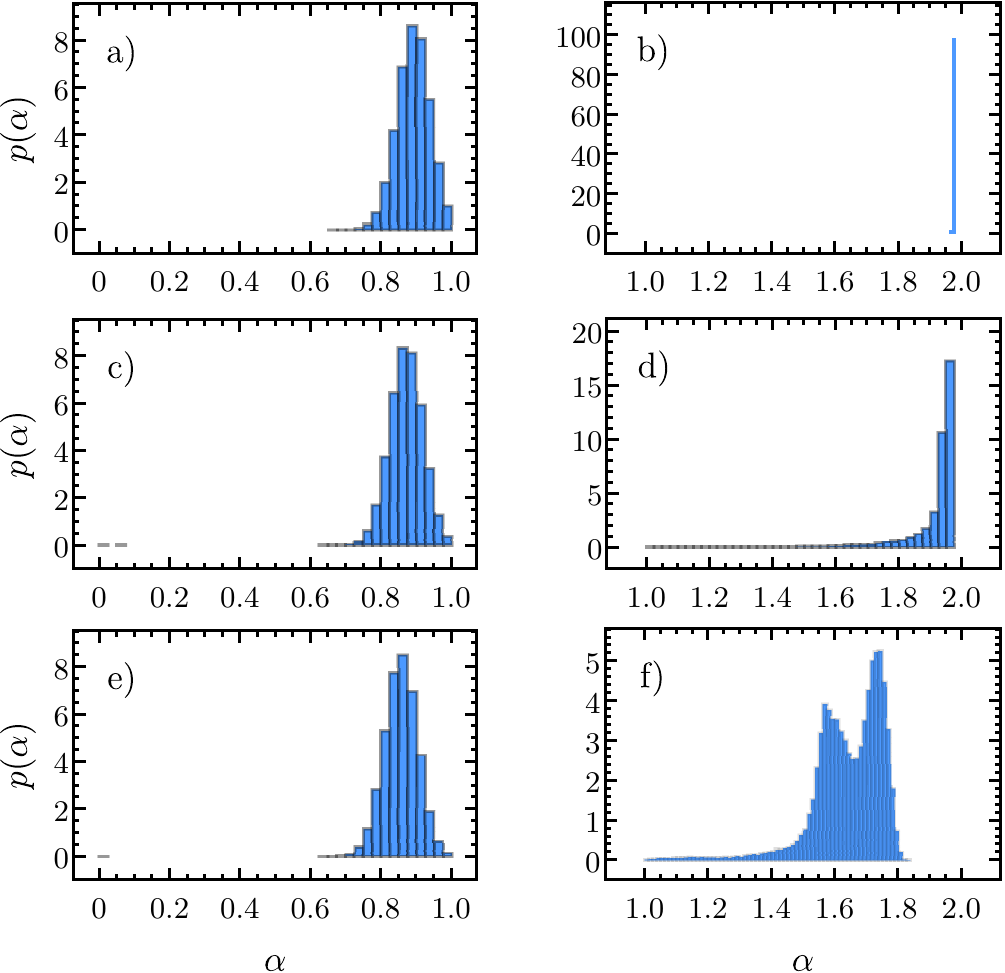} \\
    \caption{The same as in Fig.~\ref{fig:tamsdalphahist} for spherocylinders.}
    \label{fig:tamsdalphahist-sp}
\end{figure}

\begin{figure}[!h]
    \centering
    \includegraphics[width=0.9\columnwidth]{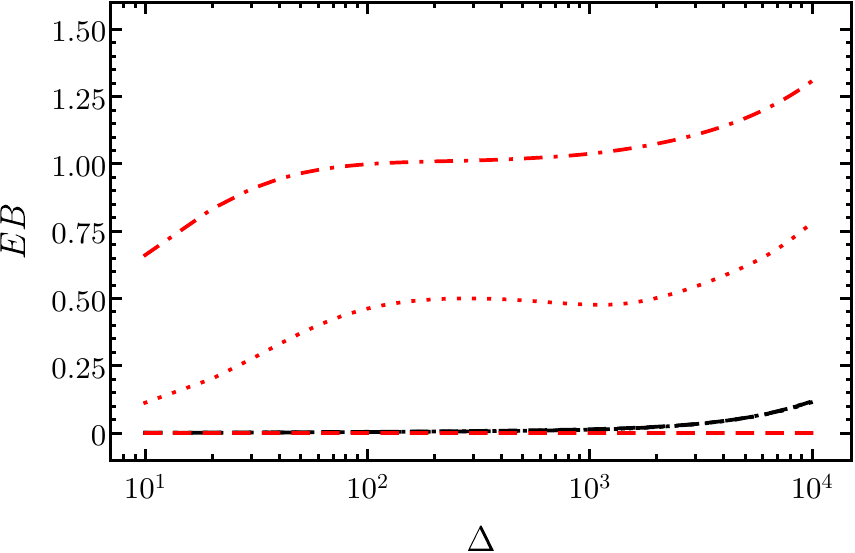} \\
    \caption{The same as in Fig.~\ref{fig:eb} for spherocylinders.}
    \label{fig:eb-sp}
\end{figure}

\bdt{
The main setup under study is presented in Fig.~\ref{fig:environment}.
The crowded environment which is well visible in Figs.~\ref{fig:environment} and~\ref{fig:2dvisited} is built by fibrinogen particles.
Fibrinogen molecules are made up of numerous globules that make them concave.
This allows diffusing molecules to accumulate in the cavities of the molecules.
The concavity can also lead to trapping in fully ordered setups, see top panel of Fig.~\ref{fig:2dvisited}.
In order to assess the role played by concavity, we have repeated studies from the main part of the manuscript for spherocylinders which are convex.
Subsequent figures
\ref{fig:2dvisited-sp},
\ref{fig:tamsd-sp},
\ref{fig:tamsdalphahist-sp},
\ref{fig:eb-sp},
are counterparts of Figs.
\ref{fig:2dvisited},
\ref{fig:tamsd},
\ref{fig:tamsdalphahist},
\ref{fig:eb}
from Sec.~\ref{sec:results}.
Moreover Tab.~\ref{tab:alpha} is supplemented by Tab~\ref{tab:alpha-sp}.
For compatibility with the system studied earlier, spherocylinders have the same length and area as fibrinogen particles.
Replacement of fibrinogen molecules with spherocylinders leads to quantitative changes while the qualitative properties of the model are quite well preserved,  indicating that the drift is the main factor responsible for system properties (including ergodicity breaking).
}

\bdt{
The biggest quantitative difference between both models is recorded for $\sigma(\theta)=0^\circ$, where $\theta$ is the angle between the $x$-axis and the long axis of the molecule, as for such a value the variance of the orientational angle spherocylinders are not capable of producing trapping events, see top panels of Figs.~\ref{fig:2dvisited} and~\ref{fig:2dvisited-sp}.
Furthermore, due to absence of shape-induced traps the spread of individual time-averaged mean square displacements is decreased, especially for $\sigma(\theta)=0^\circ$, see top right panels of Fig.~\ref{fig:tamsd} and~\ref{fig:tamsd-sp}, resulting in narrower histograms of recorded exponents $\alpha$, see Figs.~\ref{fig:tamsdalphahist} and~\ref{fig:tamsdalphahist-sp}.
Moreover, for a larger standard deviation of the angle $\theta$ between the $x$-axis and the long axis of the molecule measured  exponents are larger, see the bottom right panel of Fig.~\ref{fig:tamsdalphahist-sp} and Tab.~\ref{tab:alpha-sp}.
}

\bdt{
The change in the shape of obstacles significantly modifies the value of the ergodicity breaking parameter for $\sigma(\theta)=0^\circ$.
For spherocylinders there are no concave parts of the molecules, consequently for the ordered system there are not traps, the motion is almost ballistic, see Tab~\ref{tab:alpha-sp}, and $EB=0$, see Fig.~\ref{fig:eb-sp}.
In the remaining cases of $\sigma(\theta) \in \{20^\circ,180^\circ \}$ for fibrinogen and spherocylinders, the action of the drift results in weaker ergodicity breaking than for the environment built out of fibrinogen molecules, cf., Fig.~\ref{fig:eb} and~\ref{fig:eb-sp}.
}


\begin{table}[!h]
    \centering
    \begin{tabular}{c||c|c||c|c}
    & \multicolumn{2}{c||}{$\langle \overline{\delta^2} \rangle$} & \multicolumn{2}{c}{$\langle \Delta \bm{r}^2 \rangle$}  \\
    $\sigma(\theta)$ & $\;\Lambda=0\;$ &  $\Lambda=1$ & $\;\Lambda=0\;$ &  $\Lambda=1$ \\ \hline\hline
    $0^\circ$ & 0.89 & 1.97 & 0.89 & 1.97 \\ \hline
    $20^\circ$ & 0.87   & 1.95 & 0.87 & 1.91 \\ \hline
    $180^\circ$ & 0.86 & 1.65 & 0.86 & 1.59 \\
    \end{tabular}
    \caption{The same as in Tab.~\ref{tab:alpha} for spherocylinders.}
    \label{tab:alpha-sp}
\end{table}

%
%
\section*{Data availability}
The data that support the findings of this study are available from the corresponding author (PK) upon reasonable request.

%
%

\def\url#1{}

\end{document}